\documentclass[11pt,reqno]{amsart}

\usepackage{amsmath,amssymb,amsthm, amsfonts,xcolor}

\usepackage[colorlinks=true]{hyperref}

\usepackage[margin = 1in]{geometry}
\usepackage{graphicx}
\usepackage{float}
\usepackage{subfigure}
\usepackage{graphicx}
 \usepackage{mathptmx}

\newtheorem{conjecture}{Conjecture}
\newtheorem{lemma}{Lemma}[section]
\newtheorem{theorem}[lemma]{Theorem}

\theoremstyle{remark}
\newtheorem{remark}{Remark}
\theoremstyle{definition}

\title{Nature vs. Nurture: Dynamical Evolution in Disordered Ising Ferromagnets}

\author{Lily Z.\ Wang}
\address{L.\ Z.\ Wang \hfill \break
Center for Applied Math \\ Cornell University\\
Ithaca, NY}
\email{zw477@cornell.edu}

\author{Reza Gheissari}
\address{R.\ Gheissari\hfill\break
Courant Institute\\ New York University\\
New York, NY.}
\email{reza@cims.nyu.edu}

\author{Charles M.\ Newman}
\address{C.\ M.\ Newman\hfill\break
Courant Institute\\ New York University\\
New York, NY; NYU-ECNU Institute of Mathematical Sciences at NYU Shanghai, Shanghai, China.}
\email{newman@cims.nyu.edu}

\author{Daniel L.\ Stein}
\address{D.\ L.\ Stein \hfill \break Department of Physics and Courant Institute, New York University, New York, NY, USA; NYU-ECNU Institutes of Physics and Mathematical Sciences at NYU Shanghai, Shanghai, China; Santa Fe Institute, Santa Fe, NM, USA.}
\email{daniel.stein@nyu.edu}

\begin{document}
	
\maketitle	

\begin{abstract}
We study the predictability of zero-temperature Glauber dynamics  in various models of disordered ferromagnets. This is analyzed using two independent dynamical realizations with the
same random initialization (called twins). We derive, theoretically and numerically, trajectories for the evolution of the normalized magnetization and twin overlap as the system size tends to infinity. The systems we treat include mean-field ferromagnets with light-tailed and heavy-tailed coupling distributions, as well as highly-disordered models with a variety of other geometries. In the mean-field setting with light-tailed couplings, the disorder averages out and the limiting trajectories of the magnetization and twin overlap match those of the homogenous Curie--Weiss model. On the other hand, when the coupling distribution has heavy tails, or the geometry changes, the effect of the disorder persists in the thermodynamic limit. Nonetheless, qualitatively all such random ferromagnets share a similar time evolution for their twin overlap, wherein the two twins initially decorrelate, before either partially or fully converging back together due to the ferromagnetic drift.
\end{abstract}

\section{Introduction}

The dynamical evolution of a discrete spin system quenched from high
to low temperature remains an area of active investigation (for a
review, see~\cite{Bray94}).  In this paper we continue the study of
the nonequilibrium evolution of an instantaneous quench from infinite
to zero temperature. In such a situation, the initial spin
configuration for an Ising system at time zero is i.i.d.~Bernoulli
with $p=1/2$; i.e., every spin is $\pm 1$ with equal probability,
independently of all the others. The subsequent time evolution is
taken to be the zero-temperature limit of Glauber dynamics, to be
described below.

The behavior of the subsequent dynamical evolution has been studied
from a variety of perspectives addressing different questions that
naturally arise. Here we focus on the problem of \emph{predictability}: how
does the information contained in the initial state decay with time?
Equivalently, how much of the information contained in the
configuration at time $t$ depends on the initial configuration and how
much depends on the dynamical realization (i.e., the order in which
spins are updated and, where zero-energy flips are possible, the outcome of tie-breaking coin flips)?  We have referred to
this colloquially as the ``nature vs.~nurture'' problem~\cite{SN14,YMNS13},
where nature refers to the information contained in the initial
configuration (and quenched random couplings if the system is
disordered), while nurture refers to the realization of the dynamical
evolution of the system.

In a series of papers~\cite{NNS00,YMNS13,YGMNS17,GNS18}, this problem
was studied for systems with homogenous and random couplings (and in the latter
case, for both random ferromagnets and spin glasses), and for both
short-range interactions in finite dimensions and mean-field models with infinite-range
interactions. In this paper we focus on disordered Ising ferromagnets, primarily the mean-field situation which we call \emph{disordered Curie--Weiss models}. Here, the ferromagnetic nature of the system induces a drift towards one of the two homogeneous ground states of the model; this natural drift competes with the ``glassy" random dynamical evolution that can arise from locally stable (e.g., 1-spin-flip-stable) states, and can be observed through the predictability framework described above. We analyze the limiting trajectories as functions of time (suitably rescaled) of two observables: the absolute magnetization~$|M(t)|$ and the twin overlap~$q_D(t)$.

For the classical Curie-Weiss ferromagnet, it is simple to see that as $N\to\infty$ the final state of the system is completely determined
by the initial state; the dynamics plays no role in determining the
final state.\footnote{In fact, this statement is true for all odd finite~$N$. For even~$N$, it fails only in the case where the initial state comprises exactly half plus spins and half minus spins. The probability of realizing this initial state falls to zero as $N^{-1/2}$ for large~$N$.}  In this system the dynamics consist only of minus spins flipping to plus (or vice-versa), and therefore the twin overlap 
evolution admits an exact description by a birth and death chain (Section~\ref{sec:theory}). 

For disordered mean-field models (with, say, coupling distribution having a finite exponential moment), one expects a similar behavior to hold as $N\to\infty$, though proving this becomes substantially more difficult.
 In~\cite{GNS18}, this outcome was proven for the randomly diluted (or Bernoulli disordered)  Curie-Weiss model (i.e., a model in which between any two sites, a coupling of magnitude~1 is present with probability $p\in (0,1)$ and absent with probability $1-p$) in the following asymptotic sense: as $N\to\infty$, with high probability over the random couplings, the initial state, and the dynamical realization, the dynamics ends up in the homogenous (all-plus or all-minus) state corresponding to the sign of its magnetization at time zero. We strongly expect the same result to be true for the disordered
Curie-Weiss model on the complete graph (e.g., with a half-normal
distribution for the couplings), but this has not yet been proved: see Conjecture~\ref{conj:pos-field} below. (For the sake of comparison, if the dilution parameter $p$ is going to $0$ as $N\to\infty$ rapidly enough, say in the sparse regime of $p = O(1/N)$, the behavior is drastically different and was studied in~\cite{Hag}.)

A related setup in the physics literature that has been extensively studied is that of a $p$-spin glass spiked by a Curie--Weiss interaction term $\lambda \sum_{i,j} \sigma_i \sigma_j$; these are sometimes called $2+p$ spin models~\cite{CriLeu,CriLeu2,GiSh00}.  Namely, when the corresponding spin glass model has couplings with bounded support, our models fit into that setup with hard ($\pm 1$) spins. It is natural in these models to analyze the magnetization of either the Gibbs measure, or as a function of time with respect to some Langevin/gradient descent algorithm, as the spike strength $\lambda$ varies. In the spherical setting, the structure of local minima  and the behavior of \emph{zero-temperature} {gradient descent algorithms} have seen much recent attention~\cite{BMMN17,BBCR18,MR14,BAGJ18}.

In this paper, we extend the study of zero-temperature dynamics for random ferromagnets in several
directions: we present analytical results for  ``light-tailed models''
(i.e., models in which the coupling distribution falls off fast---e.g., exponentially) assuming the validity of Conjecture~\ref{conj:pos-field}; in particular, we
solve for the limiting time evolution of both the absolute
magnetization $|M(t)|$ and the twin overlap $q_D(t)$ (to be defined below), which
measures the degree to which the initial state and dynamics
respectively determine the later state of the system. We then
present numerical simulations that support these results and suggest that the limiting trajectories are universal (do not depend on 
the choice of light-tailed coupling distribution).

We also present numerical results for heavy-tailed distributions (with no mean or variance), which display very  different
$t\to\infty$ behavior, consistent with earlier rigorous results from Sec.~3.1 of~\cite{GNS18}.  It was proven in~\cite{GNS18} that systems with heavy tails (say, with undefined mean) typically absorb into configurations with macroscopic but not full magnetization. The numerical results give a more complete picture of the approach to absorption in random mean-field ferromagnets and their dependence on the coupling distribution: we also find a general profile to $q_D(t)$ that  independent dynamical evolutions of the random ferromagnets (whether light tailed or heavy tailed)  rapidly decorrelate from overlap 1 to some twin overlap $q_{\textsc{min}} \in (0,1)$ before partially merging on the way to some limiting $\lim_{t\to\infty} q_D(t)>q_{\textsc{min}}$.

In addition, we study the opposite extreme of a one-dimensional chain on $N$ vertices, whose time evolution can be solved analytically, and
we compare the qualitative behavior of this model with that of the mean-field models. The 1D~chain is a simple example of highly disordered random ferromagnets with a particular tree-like structure~\cite{NS96a,NN00},
which is also amenable to analytical solution, and for completeness (and as a starting point for possible future work) we present those results as well.

The paper is organized as follows. In Sec.~\ref{sec:prelims}, we define the various models we consider as well as their zero-temperature Glauber dynamics.  In Sec.~\ref{sec:theory} we derive analytical expressions that are valid after
very short times (to be specified below) in mean-field models for both the magnetization and the twin overlap, which measures
the relative degree of nature vs.~nurture and is defined below. Sec.~\ref{sec:tree} contains a theoretical discussion of the time evolution of the magnetization and 
the twin overlap in both the $1D$~random ferromagnet and the ``highly disordered'' model~\cite{NS94,NS96a} on regular trees. On more general graphs, e.g., $\mathbb Z^d$, the highly disordered model generates a tree-like structure
in the dynamics, and its time evolution can be expressed as a sum over lattice animals. In Sec.~\ref{sec:numerical} we turn to our numerical results and first describe the methods used. In Sec.~\ref{subsec:light}
we present numerical results for mean-field ferromagnets with light-tailed coupling distributions (Bernoulli and half-normal). In Sec.~\ref{subsec:heavy} we do the same with heavy-tailed distributions (half-Cauchy), and in Sec.~\ref{subsec:1D-numerics} we present numerical results for the 1D random ferromagnet.

\section{Preliminaries}
\label{sec:prelims}

As indicated in the introduction, we primarily focus on disordered Curie-Weiss Ising models, i.e., mean-field Ising ferromagnets with i.i.d.~couplings~$J_{ij}$ chosen from a common (non-negative) distribution. For all of these the system Hamiltonian is
\begin{equation}
\label{eq:CW}
H=-\frac 1N \sum_{1\leq i<j \leq N} J_{ij}\sigma_i\sigma_j\, ,
\end{equation}
where $\sigma_i=\pm 1$ denotes the spin at site~$i$, with the sites~$i$ sitting at the vertices of a complete graph.
The rescaling factor $N^{-1}$ preceding the sum on the right-hand side ensures a sensible thermodynamic limit of the energy and free energy per spin. 

We study four types of models: two which fall into the category of light-tailed distributions, and two which use heavy-tailed distributions.
The light-tailed distributions studied include the \emph{randomly diluted}, or \emph{Bernoulli}, model described in the introduction; in that model, $J_{ij}\sim \mbox{Bernoulli}(p)$ for some $p\in (0,1)$---clearly this corresponds to taking the homogenous Ising model on an Erdos--Renyi random graph $\mathcal G(N,p)$. 
We also consider the \emph{half-normal} case where the common distribution of $J_{ij}$ is taken to be a 
one-sided Gaussian, in which each coupling is chosen as the absolute value of a standard Gaussian random variable (with mean zero and variance one).

The two heavy-tailed distributions include a Pareto-like model in which $J_{ij}$ has density
\begin{equation}
\label{eq:Pareto}
p_{J}(x)=\begin{cases}
\alpha x^{-(1+\alpha)} & \text{for } x\ge 1\\
0 & \text{for } x<1\\
\end{cases}\,,
\end{equation}
with $\alpha>0$, and a half-Cauchy distribution in which
\begin{equation}
\label{eq:Cauchy}
p_{J}(x)=\begin{cases}
(2/\pi) (1+x^2)^{-1} & \text{for } x\ge 0\\
0 & \text{for } x<0\\
\end{cases}\,.
\end{equation}
The choice of these distributions is fairly arbitrary, and we believe the results to be qualitatively independent of the choice of distribution (though sensitive to whether it is light-tailed or heavy-tailed). 

In Sec.~\ref{sec:tree} we also present analytical results for several short-range models with tree-like structures, including the limiting case of a 1D~chain; for these models there is an underlying graph structure $G= (V, E)$ and the Hamiltonian is given by
\begin{equation}
\label{eq:nn}
H(\sigma)=-\sum_{\{i,j\}\in E} J_{ij}\sigma_i\sigma_j\,.
\end{equation}
Of course, this geometry could be encoded into the complete graph by setting $J_{ij}$ on the complete graph to be zero if $ij\notin E$, but then the couplings would not be i.i.d..

We are interested in the Glauber dynamics of these systems in the case where the system is instantaneously quenched from infinite to zero temperature. This is modeled by the \emph{zero-temperature Glauber dynamics}, which is the following Markov chain $(\sigma(t))_t$ with state space $\{\pm 1\}^V$. Begin by choosing the initial state~$\sigma(0)$ to be a random spin configuration
in which each spin is chosen to be $+1$ or $-1$ with probability 1/2, independently of all the others, corresponding to an infinite-temperature spin configuration. Then, at each time increment $t= \frac 1N, \frac 2N, ...$ select a site amongst $\{1,...,N\}$ uniformly at random. When the site~$i$ is selected, the energy change $\Delta E_i$ associated with flipping~$\sigma_i$ is computed: $\Delta E_i=\Delta H$ when $\sigma_i \rightarrow -\sigma_i$ and all other spins remain fixed. If the energy decreases ($\Delta E_i < 0$) as a result of the flip, the flip is carried out. 
If the energy increases ($\Delta E_i > 0$), the flip is not accepted. If the energy remains the same ($\Delta E_i = 0$), 
a flip is carried out with probability 1/2. This last ``tie-breaking'' rule is relevant only for models where the distribution of $J$ has atoms (such as the randomly-diluted Curie--Weiss model). For disordered models with continuous coupling distributions, such as the half-normal or either of our heavy-tailed distributions, this possibility almost surely never arises. In all cases the dynamics are run until the system reaches an absorbing state that is stable against all single spin flips. 

The time rescaling here by $1/N$ as compared to traditional discrete-time increments is so that an $N\to\infty$ limit yields a continuous-time chain with rate-1 Poisson clocks without any additional rescaling.
One \emph{sweep} consists of $N$~time increments $n+\frac{1}{N},n+\frac{2}{N},...,n+1$ between two integer times. 

We are interested in both the time dependence and final values of both the absolute value of the magnetization and the ``twin overlap''. For an Ising spin system on $N$ sites evolving at zero temperature according to the above dynamics, we denote the magnetization per spin by 
\begin{align}\label{eq:def-M}
M_N(t) = \frac 1N \sum_{i=1}^{N} \sigma_i(t)\,.
\end{align}
The \emph{dynamical twin overlap} was introduced in~\cite{NNS00} and studied further in e.g.,~\cite{YMNS13,YGMNS17}; we briefly recount its definition here.
Two Ising~systems $\sigma(t),\sigma'(t)$ (regarded as identical twins) with $N$~spins are prepared with the same initialization~$\sigma(0)$, 
and are then allowed to evolve using independent realizations of the zero-temperature Glauber dynamics described above. The spin
overlap between the resulting copies at time $t$ is  
\begin{align}\label{eq:def-q}
q_N(t)=\frac{1}{N}\sum_{i=1}^{N}\sigma_{i}(t)\sigma_{i}'(t)\,.
\end{align}
In the above and throughout the paper, we suppress the dependence on the coupling realization $J$ and initial configuration $\sigma(0)$ as they will be fixed by the context. 
Clearly $q_N(0)=1$ for any~$N$.  In systems where the $N\to\infty$ limit of $q_N(t)$ is sensible, such as nearest-neighbor models
on $\mathbb Z^d$ (where this is easily seen using translation invariance and ergodicity reasoning~\cite{NNS00}), we can also define the final overlap as $q_D(t):=\lim_{N\to\infty}q_N(t).$\footnote{Here the ``D" stands for ``dynamical'', not dimension.} For instance, if we let $\mathbb E_{J}$, $\mathbb E_{\sigma(0)}$, and $\mathbb E_{\omega}$ denote the expectations with respect to the couplings, initialization, and dynamics respectively, we have on $\mathbb Z^d$, $q_D(t) = \mathbb E_{J} \mathbb E_{\sigma(0)} \big[ (\mathbb E_{\omega} [\sigma_0(t)])^2\big]$. 

In our numerical simulations, a ``trial'' consists of a pair of dynamical runs with a single set of twins, with 50 runs for every dynamical realization of the couplings, and 50 different coupling realizations. The resulting twin overlaps at each time are therefore averaged over 2500 such trials to arrive at the value of $q_N(t)$ for that particular model.

\section{Evolutions of $M(t)$ and $q_D(t)$ in the dilute Curie--Weiss model}
\label{sec:theory}

In this section, we consider the evolution of a pair of zero-temperature
chains independently evolving from the a configuration in which
every site $i$ has a positive effective field:  
\begin{align*}
m_i(t) := \sum_{j} J_{ij} \sigma_j(t)\,,
\end{align*}
 (and therefore, the dynamics thenceforth only
consists of flipping $-$'s to $+$'s when they update). The true
behavior of disordered mean-field ferromagnets should be, with high
probability, governed by this chain after a short $o(1)$ burn-in time.
Namely, recall a theorem of~\cite{GNS18} proving this is true in the case of the randomly-diluted Curie--Weiss model with parameter $p\in (0,1)$ (corresponding to a disordered mean-field ferromagnet with $\mbox{Bernoulli}(p)$ couplings):

\begin{theorem}[{\cite[Theorem 1]{GNS18}}]\label{thm:pos-field}
Consider zero-temperature dynamics $\sigma(t)$, for the randomly-diluted model with $p\in (0,1]$. For all $\delta>0$ sufficiently small, for every $\sigma(0)$ with $M_N(0)\geq N^{-\frac 12 -\delta}$
\begin{align*}
\lim_{N\to\infty} \mathbb P_{J} \times \mathbb P_{\omega} \big(m_i (N^{-\frac 12 + 3\delta})> 0\, \mbox{ for all $i$}\big) =1\,.
\end{align*}
\end{theorem}
Notice that a random initialization has $M_N(0)$ that is of order $N^{-\frac 12}$, so that indeed the above, or its spin-flipped version, applies with high probability to a random initial configuration. 

\begin{conjecture}\label{conj:pos-field}
The conclusion of Theorem~\ref{thm:pos-field} holds for a disordered Curie-Weiss model with i.i.d.\ non-negative couplings that are sub-exponential.
\end{conjecture}

After this $O(N^{-\frac 12 +3\delta})$ burn-in time,  if indeed every spin feels a positive effective field, the evolution of $M_N(t)$, as well as the evolution of the twin overlap $q_N(t)$,
are given exactly by simple birth-and-death chain computations. We give this
theoretical solution via the transfer matrix of a projection of the original Markov chain in what follows, and show that it matches the
experimental results in Section~\ref{sec:numerics}. 

The theoretical predictions here also match well with the numerical results for the 
half-normal distribution (Sec.~\ref{subsec:light}), as well as the  uniform distribution on $[0,1]$ (not-presented in this paper). This suggests a universality for the zero-temperature dynamics in the observables $M(t),q(t)$ when the
disorder is light-tailed. Namely, after a short period of time during
which the magnetization performs a biased random walk and escapes the
set of sites with ``typical" $O(\sqrt N)$ magnetization, the drift
becomes so strong that the evolution of the dynamics becomes
essentially deterministic, and independent of the disorder.

We emphasize that this behavior differs substantially for heavy-tailed distributions; see Section~\ref{subsec:heavy}.  

\subsection{A simplified Markov Chain}\label{sec:simplified-chain}
As after some $o(1)$ time, and $o(N)$ many spin flips, the zero-temperature Glauber dynamics chains are in configurations such that every spin flip is from minus to plus (or vice versa), let us begin with the following simplified Markov chain. 
Consider a discrete-time Markov chain $(\sigma(t),\sigma'(t))$ in $\{\pm 1\}^N\times \{\pm 1\}^N$ that at each integer time independently selects two sites $v,v'$ uniformly at random amongst $\{1,...,N\}$, and if its state $\sigma_{v}(t)$ (resp., $\sigma'_{v'}(t)$) is minus, flips it to plus. Notice that this exactly describes the evolution of the \emph{homogenous} Curie--Weiss model, where all $J_{xy} = 1$, as long as the configuration starts with an imbalance of $+$'s and $-$'s. 

We can project this Markov chain onto four variables $$\mathbf N= (N_{++},
N_{+-}, N_{-+}, N_{--})$$ that count the number of sites that are plus in both chains, plus in the first and minus in the second, etc..., with the constraint that $N_{++}+
N_{+-} + N_{-+}+ N_{--}= N$. 
The projected chain then has the following transition probabilities:
\begin{align*}
P(N_{++}^{t}, \cdot) & = \frac{N_{+-}^{t}}{2N} + \frac{N_{-+}^{t}}{2N} \delta_{N_{++}^t+1}\,, \qquad\qquad
& P(N_{-+}^t, \cdot )  = \frac{N_{--}^t}{2N} \delta_{N_{-+}^t+1} + \frac{N_{-+}^t}{2N} \delta_{N_{-+}^t-1}\,, \\
P(N_{--}^t, \cdot ) & = \frac{N_{--}^t}{N} \delta_{N_{--}^t-1}\,, \qquad \qquad 
& P(N_{+-}^t, \cdot ) = \frac{N_{--}^t}{2N}\delta_{N_{+-}^t +1} + \frac{N_{+-}^t}{2N}\delta_{N_{+-}^t -1} \,.
\end{align*}
From \emph{any} initial state, this chain absorbs into
$N_{++}^T=N$ in a time $T$ that is $O(N \log N)$ with high probability. Therefore, by standard martingale arguments applied to each of $\frac{1}{N}(N_{++}, N_{+-}, N_{-+}, N_{--})$---at each time step at most one spin flip occurs so that their martingale increments are bounded by $\frac{1}{N}$ and we can apply, e.g., Azuma's inequality---the trajectories of the four components of $\frac{\mathbf N^t}N$ concentrate around their means with maximal fluctuations of order $O(N^{-\frac 12+o(1)})$.  Hence we can move to the process
governing the evolution of $\mathbb E[N_{++}^t]$ etc., denoting these means
by $\bar {\mathbf N}^t = (\bar N^t_{++}, \bar N^t_{+-}, \bar N^t_{-+}, \bar
N^{t}_{--})$.
With the transition probabilities described above, we have $\bar{\mathbf N}^t = \mathbf A^t \mathbf N^0$ for every $t\geq 0$, where 
\begin{align*}
\mathbf A = \left(\begin{array}{cccc}1 & \frac{1}{2N} & \frac{1}{2N} & 0
  \\ 0 & 1-\frac{1}{2N} & 0 & \frac{1}{2N} \\ 0 & 0 & 1-\frac{1}{2N} &
  \frac{1}{2N} \\ 0 & 0 & 0 & 1-\frac{1}{N} \end{array} \right)\,.
\end{align*}
By explicit computation, we see that $\mathbf A$ can be diagonalized with
eigenvalues $\lambda= 1,1-\frac{1}{2N}, 1-\frac{1}{2N}, 1-\frac 1N$ and we
can write $\mathbf A^t$ explicitly as the upper-triangular matrix 
\begin{align}\label{eq:transition-matrix-t}
\mathbf A^t = \left(\begin{array}{cccc}1 & 1-a^t & 1-a^t & 1-2a^t+b^t \\ 0 & a^t & 0 & a^t-b^t \\ 0 & 0 & a^t & a^t-b^t \\ 0 & 0 & 0 & b^t \end{array} \right)
\end{align}
where $a= (1-\frac{1}{2N})^t$ and $b= (1-\frac 1N)^t$. 

\subsection{The disordered Curie--Weiss model}
From the earlier discussion and Theorem~\ref{thm:pos-field}, it follows that for the randomly-diluted model, after a short $O(N^{-\frac 12 +\delta})$ burn-in time, with high probability the evolution of $q_N(t)$ and $M_N(t)$ are governed by the chain analyzed in Section~\ref{sec:simplified-chain} (up to the described time-change of $t\mapsto tN$).  
 Based on the ansatz (supported by the numerics in \S\ref{sec:results}), that the
 light-tailed models should all exhibit the same qualitative zero-temperature
 behavior, it is reasonable to assume that this limiting trajectory
 of $(q_D(tN))_{t\geq 0}$ is accurate also for disordered
 Curie--Weiss models with more general light-tailed distributions.
 
Now let $\mathbf N^t$ be defined as the projection of two independent copies of zero-temperature Glauber dynamics, so that e.g., $$N_{+-}^t= \sum_i \mathbf 1\{\sigma_i(t) = +1, \sigma'_i(t) = -1\}\,.$$
With probability tending to $1$ as $N\to\infty$, starting from an infinite-temperature quench,
either $N_{++}^0 \geq N^{\frac 12 -\delta}$ or $N_{--}^0 \geq N^{\frac 12 -
  \delta}$. By spin-flip symmetry we consider the former. By Theorem~\ref{thm:pos-field}, for small enough $\delta>0$,
after some time $T := N^{-\frac 12 +3\delta}$ every site feels a positive
effective field, at which point the dynamics of the two independently evolving twins are
completely determined by the Markov chains defined in \S\ref{sec:simplified-chain}. Since $T= o(N)$ and at most one spin flips at each time step, 
it suffices to consider the chain defined in \S\ref{sec:simplified-chain} initialized from
\begin{align*}
\mathbf N^0 = \Big(\frac{N}2 +o(N), o(N), o(N), \frac N2 + o(N)\Big)\,.
\end{align*}

By the time scaling used in the different time increments for the zero-temperature Glauber dynamics chain we defined and this simplified alternate Markov chain, every time increment of one in the former corresponds to $2N$ steps in the latter, so that  
\begin{align}\label{eq:mean-q-D}
\mathbb E[q_N(t)] = \frac 1N (\bar{N}_{++}^{2tN} + \bar{N}_{--}^{2tN} - \bar{N}_{+-}^{2tN} - \bar{N}_{-+}^{2tN})+{o(1)}\,.
\end{align} 
Plugging the initial data into~\eqref{eq:mean-q-D} and applying the transition matrix~\eqref{eq:transition-matrix-t}, we obtain 
\begin{align*}
\mathbb E[q_N(t)] = 1-a^{2tN} + b^{2tN} -(a^{2tN}-b^{2tN}) + o(1)= 1+2b^{2tN} - 2a^{2tN}+o(1)\,.
\end{align*}

Sending $N\to\infty$ and plugging in for $a,b$, we observe that for every fixed $t\geq 0$, 
\begin{align}
\label{eq:qteq}
\lim_{N\to\infty} \mathbb E[q_N(t)] = 1+2e^{-2t}-2e^{-t}\,.
\end{align}
As before, by Azuma's inequality  and the fact that the chain absorbs to $N_{++}= N$ after $O(N\log N)$ single spin-flips with high probability, for every $\delta>0$,
\begin{align}\label{eq:concentration}
\lim_{N\to\infty}\mathbb P\Big(\sup_{s\geq 0} |q_N(s) - \mathbb E[q_N(s)] | \geq N^{-\frac 12 +\delta}\Big) =0\,.
\end{align}
As a consequence of~\eqref{eq:concentration}, for every $\delta>0$, with high probability under $\mathbb P_{J}\times \mathbb P_{\sigma(0)}$, 
\begin{align}\label{eq:result1}
\lim_{N\to\infty} \mathbb P_\omega \left(\sup_{s\geq 0} \big |q_N(s)- (1+2e^{-2s}- 2e^{-s})\big | \geq N^{-\frac 12 +\delta}\right) = 0\,.
\end{align}
Notice that this matches with Figures~\ref{fig:Bernoulliqt}, and
the minimum value of $q_D(s)$ is $\frac 12$.

By analogous (simplified) reasoning, we also see that the evolution of the mean magnetization is described by $1-e^{-2s}$, and as a consequence,  for every $\delta>0$, with high probability under $\mathbb P_{J} \times \mathbb P_{\sigma(0)}$, 
\begin{align}\label{eq:mag-theory}
\lim_{N\to\infty} \mathbb P_\omega \Big(\sup_{s\geq 0} \big||M_N(s)| - (1-e^{-2s})\big|\geq N^{-\frac{1}2 +\delta}\Big) =0\,.
\end{align}

\section{Evolution of $q(t)$ in a more general setting}
\label{sec:tree} 

Using the above calculations as a springboard, we are able to give a more general expression 
for the evolution of $q_D(t)$ for a certain class of random ferromagnets 
on more general graphs than the complete graph. In a ``highly disordered" 
setting, which we will describe below, the evolution of $q_N(t)$ is dictated 
by the evolution of a pair of independent random walks on certain tree-like 
graphs, dictated by the coupling disorder. We emphasize that  
mean-field models with light-tailed couplings are far from ``highly disordered", but the highly-disordered picture may give  a good heuristic for the behavior when the couplings are heavy-tailed 
and therefore live on all scales (Section~\ref{subsec:heavy}). 

The general highly-disordered framework also includes the nearest-neighbor graph in one-dimension with any continuous coupling distribution. Let us begin with this specific model, for which we have matching numerics in Section~\ref{subsec:1D}. 
As in the rest of the paper, we will restrict attention to disordered ferromagnets,
where the individual coupling variables $J_{xy}$ 
or $J_e$ for (undirected) edges $e=\{x,y\}$
are i.i.d. with common (continuous) distribution $\nu$ supported on $(0,\infty)$.
We note though that specifically in this highly-disordered setting, our results are essentially unchanged without the ferromagnetic
restriction so that the distribution can be supported also on negative
values. This includes the spin glass case, where $\nu$ is an even distribution
on $(-\infty,\infty)$. (C.f.\ the non-highly disordered case, where the spin-glass models e.g., the traditional Sherrington--Kirkpatrick model, display dramatically different behavior from the light-tailed random ferromagnet~\cite{GNS18,YGMNS17}.)

\subsection{Evolution of $q_D(t)$ in the 1D setting} 
\label{subsec:1D}

In a finite segment of the one-dimensional lattice, say 
$G_N = ({ V}_N, { E}_N)$ with $N$ vertices or sites,
${V}_N = \mathbb Z/n\mathbb Z$, and $N$ nearest-neighbor edges 
${E}_N = \{\{0,1\}, \{1,2\}, \dots ,\{N-1,0\}\}$, we may use discrete
time updating at times $t = \frac{1}{N},\frac{2}{N}, \dots$ when we choose a site $x$ uniformly
at random and then update the $\pm 1$ value $\sigma_x(t)$ by flipping
the value if that lowers the energy. Equivalently, we may define $z(x)$
to be the neighbor $y$ of $x$ with the largest magnitude coupling among the
two neighbors: $|J_{x,z}| = max \{|J_{x,x-1}|, |J_{x,x+1}|\}$ and update 
$\sigma_x(t)$ to equal $\sigma_{z(x)} (t-1/N)$. 

We study the evolution in time of two identical twins, $\sigma(t)$ and $\sigma'(t)$, initialized by assigning  to $\sigma_{x}(0)=\sigma'_x(0)$ $\pm 1$-spins with i.i.d.\ fair coin flips, evolving under independent evolutions (i.e.,
with independent choices for the two twins of which site updates at each time step). Recall the notations $\mathbb P_{J}, \mathbb P_{\sigma(0)}$ and $\mathbb P_{\omega}$ for the randomness from couplings, initialization, and dynamics, and when we combine subscripts, e.g., $\mathbb P_{J, \sigma(0)}$, we mean the product measure. 

Under this process, the twin overlap $q_N(t)$ has mean
\begin{align}\label{eq:overlapformula-finite-N}
\frac 1N \sum_x \mathbb E_{J,\sigma(0), \omega} [\sigma_x(t) \sigma'_x (t)]
= \frac 1N \sum_x \mathbb E_{J, \sigma(0)} [\mathbb E_{\omega} (\sigma_x(t))]^2 \, .
\end{align}

In the limit as $N \to \infty$, where we always identify the vertex $0$ in $\mathbb Z/n\mathbb Z$ with the origin of $\mathbb Z$ and shift the torus to be centered about its origin, the limiting graph has, 
as $N \to \infty$, the vertex set $\mathbb Z$. By defining an analogous dynamics on $\mathbb Z$ via i.i.d.\ Poisson clocks at each vertex (as opposed to discrete time steps), one obtains a valid limiting zero-temperature Markov process on $\mathbb Z$ for which, by translation invariance and ergodicity, one has 
\begin{align}
\label{eq:overlapformula}
q_D(t) = \mathbb E_{J,\sigma(0),\omega} [\sigma_0(t) \sigma'_0 (t)] = \mathbb E_{J,\sigma(0)} \big[\mathbb E_{\omega} [\sigma_0(t)]^2\big] \,.
\end{align}
where now the expectations are with respect to the couplings, initialization, and dynamics on all of $\mathbb Z$. Because this limit is nicely defined, let us work directly with this limiting quantity.

There are two main ingredients needed to obtain exact expressions for $q_D(t)$.
The first is the structure of the random directed path starting from the origin~$0$ 
to $z(0)$ to $z(z(0))$ to \dots (as dictated by the random couplings), and the second is a geometric
analysis of the dynamic processes $\sigma_0(t), \sigma'_0(t)$ in terms of (backward in time)
random walks along these directed paths.

For the nearest neighbor graph on $\mathbb Z$ with continuous coupling distribution (as well as for the highly disordered
models we will consider on other graphs) the random directed graph with directed
edges $(x, z(x))$ will have the following two properties (with probability one):

(A). The only possible directed cycles, $(x_0,x_1), (x_1,x_2), \dots, (x_{n-1},x_n)$,
with $x_k = z(x_{k-1}) \forall k$ and with $x_n = x_0$, are those with $n=2$ --- i.e.,
$(x_0,x_1), (x_1,x_0)$ with $x_1 = z(x_0)$ and $x_0 = z(x_1)$. In the language of~\cite{NNS00} that makes the undirected edge $(x_0, x_1)$ a {\it bully bond\/}
so that when either of the two vertices updates, it does so by making that
particular bond satisfied (i.e., in the ferromagnetic case, makes the two spin
values on the endpoints of that bond agree with each other) regardless of the state
of other spins. 

(B). For any $x_0$, either $(x_0, z(x_0))$ is itself a bully bond or else there is
a minimum $m \geq 1$ such that the directed path starting from $x_0$ is of the form
$(x_0,x_1), (x_1,x_2), \dots, (x_{m-1},x_m)$ with $(x_m,z(x_m))$ a bully bond. Define
the random variable $M(x_0)$ to be~$0$ if $(x_0, z(x_0))$ is a bully bond, and
otherwise to be this minimum~$m$. This $M(x_0)$ is the number of steps from $x_0$ to
its associated bully bond. 

In the nearest-neighbor graph on $\mathbb Z$ the distribution of ${\tilde M} := M(0)$ 
is easy to determine and is given by
\begin{align}
\mathbb P_{J} (\tilde M \geq k) = \frac{2}{(k+2)!} \qquad \mbox{and}\qquad
\mathbb P_{J} (\tilde M = k) =  \frac{2(k+2)}{(k+3)!}\,. \,
\end{align}
The corresponding distributions for other graphs in the highly
disordered limit can be more complicated, especially for the
nearest neighbor graph on $\mathbb Z ^d$  for $d \geq 2$: we give exact expressions for the complete graph in~\eqref{eq:complete-graph-tilde-m} and homogenous tree in~\eqref{eq:homogenous-tree-tilde-m}. 

The second ingredient we need to evaluate $q(t)$ is the backwards in time
representation of $\sigma_0(t), \sigma'_0(t)$. Since we have already let $N \to \infty$,
we may consider continuous time dynamics, where updates are implemented
according to rate one Poisson process clocks at the different sites. The clocks
are independent at different sites and also independent for the two twins.
With $x_0 =0$ and $x_j = z(x_{j-1}) \forall j \geq 1$, we have that starting
from $j= {\tilde m}$ (where ${\tilde m}$ is the distance from the origin~$0$ to
its bully bond at $\{x_{\tilde m}, x _{\tilde m +1}\}$), the $x_j$'s are just 
$x_{\tilde m}, x_{\tilde m +1}, x_{\tilde m}, x_{\tilde m +1}, \dots$ with 
$x_0, x_1, \dots, x_{\tilde m +1}$ all distinct.

Now let $W(\tau)$ for $\tau \geq 0$ be a continuous time random walk with
$W(\tau)$ successively equal to $x_0, x_1, \dots, x_j, \dots$ during the
time intervals $[0, T_1)$, $[T_1, T_1 + T_2), \dots, [T_1 +\dots T_j, T_1 + \dots T_{j+1})$
and so forth, where $T_1, T_2, \dots$ are i.i.d. exponential (mean one) random
variables. Also let $W'(\tau)$ be an independent copy of $W(\tau)$ to
be used for the second twin.  
Then for fixed disorder and fixed $\sigma(0)$, the pair $(\sigma_0(\tau), \sigma'_0(\tau))$ is
equidistributed (for $0 \leq \tau \leq t$) with the pair
$(\sigma_{W(\tau)} (t-\tau), \sigma'_{W'(\tau)} (t - \tau))$. Thus taking $\tau = t$, one
has equidistribution of $\sigma_0(t), \sigma'(t)$ with $\sigma_{W(t)}(0), \sigma_{W'(t)} (0)$,
which implies (where we use the subscript $\textsc{walks}$ to indicate the distribution or
average over the pair of random walks):
\begin{align} 
\mathbb E_{\sigma(0)} \mathbb E_{\omega} (\sigma_0(t) \sigma_0(t')) = 
\mathbb E_{\sigma(0)} \mathbb E_{\textsc{walks}} (\sigma_{W(t)}(0) \sigma_{W'(t)}(0))
= \mathbb E_{\textsc{walks}} (1_{W(t) = W'(t)} ) = \mathbb P_{\textsc{walks}} (W(t) = W'(t)),
\end{align} 
and thus 
\begin{align}
q_D(t) = \mathbb E_{J} \mathbb P _{\textsc{walks}} (W(t) = W'(t))
= \sum_{\tilde m =0}^\infty \mathbb P _{J}(\tilde M = \tilde m) 
\mathbb P _{\textsc{walks}} (W(t)= W'(t) | \tilde M = \tilde m) \, .
\end{align}

\begin{remark}
Indeed one can recover the result that $\lim_{t\to\infty} q_D(t) = \frac 12$ (proven in~\cite{NNS00}) from this equation: for each finite $\tilde m$, the probability that both walks have reached the bully bond goes to one as $t\to\infty$; after they have both reached the bully bond, as $t\to\infty$ they are equally likely to be at either of $x_{\tilde m}$ and $x_{\tilde m+1}$, independently, so that for every $\tilde m$,
\begin{align}
\lim_{t\to\infty} \mathbb P _{\textsc{walks}} (W(t)= W'(t) | \tilde M = \tilde m) = \frac 12\,.
\end{align}
More interestingly, we will see in Sec.~\ref{subsec:highly-disordered} that in arbitrary highly disordered models, the corresponding equation here is only changed in the distribution of $\tilde M$, a random variable which is still finite almost surely; as such, we deduce that for all such highly disordered models, the limit $\lim_{t\to\infty} q_D(t)$ is $\frac 12$ and the different underlying graphs only change the trajectory on the way to this long time limit. 
\end{remark}

Let $\mathcal N$ (resp., $\mathcal N'$) denote the number of steps
taken by $W(\cdot)$ (resp., $W'(\cdot)$) during the time interval
$[0,t]$. 
$\mathcal N$ is the largest $j$ such that $T_1 + \dots + T_j \leq t$ 
(and equals zero if $T_1 > t$); it has a Poisson (mean $= t$) distribution. 
This yields the following formulas:
\begin{align}
q_D(t) & = \sum_{\tilde m=0}^\infty {\mathbb P}_{J} (\tilde M = \tilde m)
\Big [\sum_{0\leq k \leq \tilde m -1} (\frac{t^k}{k!} e^{-t})^2
+ (\sum_{\substack{k \in 2\mathbb N \\ k \geq \tilde m}} \frac{t^{k}}{k!} e^{-t}   )^2
+ (\sum_{\substack{k \in 2\mathbb N + 1 \\ k\geq \tilde m}} \frac{t^{k}}{k!} e^{-t}   )^2\Big] \\
& = e^{-2t} \sum_{\tilde m=0 }^\infty {\mathbb P}_{J} (\tilde M = \tilde m)
\Big[\sum_{0\leq k \leq \tilde m -1} \frac{t^{2k}}{(k!)^2} 
+ (\sum_{\substack{k \in 2\mathbb N \\ k \geq \tilde m}} \frac{t^{k}}{k!} )^2 
+ (\sum_{\substack{k \in 2\mathbb N + 1 \\ k\geq \tilde m}} \frac{t^{k}}{k!} )^2\Big] \,. \label{eq:1D-theory}
\end{align}

\subsection{Evolution of $q_D(t)$ for highly disordered models on other graphs}\label{subsec:highly-disordered}

To briefly explain what highly disordered models are, we note that a key feature
of the one-dimensional dynamics was the existence of a unique $z(x)$ for each
vertex $x$ such that $\sigma_x$ is updated to make the bond between $x$ and $z(x)$ satisfied --- i.e.,
in the ferromagnetic setting $\sigma_x$ is updated to take on the value of $\sigma_{z(x)}$. For
other graphs this will only be the case if the (ferromagnetic) couplings $J_{xy}$
between x and all its neighboring vertices satisfy
\begin{align}
J_{x,z(x)} \, > \, \sum_{y \neq z(x)} J_{xy} \, .
\end{align}
Here the sum is over all neighboring vertices of $x$ except $z(x)$. 
For a sequence of finite graphs with $N$ vertices, such as the complete  graph
$K_N$ or finite volume portions of the hypercubic lattice ${\mathbb Z}^d$, one
can guarantee that this condition will be satisfied for each $x$ for all large $N$
(with probability one) by choosing the disorder measure $\nu =  \nu_N$ appropriately,
as discussed in~\cite{NS96a}. This can be done, for example,
by having a set of i.i.d. uniformly distributed (on $[0,1]$) random variables
$U_e$ for all the nearest-neighbor edges of ${\mathbb Z}^d$ and then letting
$J_e = \exp (-\lambda_N U_e)$ with $\lambda_N$ chosen to increase sufficiently
rapidly with $N$. In that case $z(x)$ is simply determined so that $U_{\{x,z\}}$
is larger than $U_{\{x,y\}}$ for all other neighbors $y$ of $x$. 

In this more general setting of highly disordered dynamics, the analysis leading
to a formula for the overlap $q_D(t)$ is essentially the same as for the
one-dimensional nearest-neighbor graph except that the distribution
of the random variable $\tilde M$ will depend on the graph. For example, for
the complete graph $K_N$ in the limit $N \to \infty$, one finds that
\begin{align}\label{eq:complete-graph-tilde-m}
{\mathbb P}_{J} ({\tilde M} = k)  = \, \frac {k+1}{(k+2)!}\,,
\end{align}
(though in this case, the equality of equations~\eqref{eq:overlapformula-finite-N} and~\eqref{eq:overlapformula} requires more care as there is no natural translation operator). 
For the homogeneous tree with coordination number (i.e., number of neighbors of
each vertex) $\ell$ , one finds
\begin{align}\label{eq:homogenous-tree-tilde-m}
{\mathbb P}_{J} ({\tilde M} = k) = \, \frac {\frac{1}{\ell -1} + k +1}
{(\frac{1}{\ell -1} + k+ 2) \cdot (\frac{1}{\ell -1} + k+ 1) \cdots (\frac{1}{\ell -1} + 2)} \, .
\end{align}
On the integer lattices $\mathbb Z^d$ with $d\geq 2$, the distribution of ${\tilde M}$ can be expressed as a sum over lattice animals rather than as a closed-form expression. 

\section{Numerical results}
\label{sec:results}\label{sec:numerics}

In this section we overview our main numerical results for the twin evolutions of disordered mean-field ferromagnets under zero-temperature Glauber dynamics. We first describe the numerical methods used in Section~\ref{sec:numerical}, then describe the results first for the disordered Curie--Weiss models with light-tailed distributions in Section~\ref{subsec:light}, then proceed to heavy-tailed distributions in Section~\ref{subsec:heavy}.

\subsection{Numerical methods}
\label{sec:numerical-methods}
\label{sec:numerical}

Throughout this section, time~$t$ is measured in sweeps with one sweep corresponding to $N$ distinct spin-flip attempts, as noted In Sec.~\ref{sec:prelims}. This also matches the normalization used in the theoretical section above in order to obtain meaningful $N\to\infty$ scaling limits. A single \emph{graph realization}, or just \emph{realization} consists of a sample of $\{J_{i,j}\}$. For a given realization, a single \emph{trial} consists of generating a new random initial configuration $X_0$ and using two independent dynamical runs, one for each twin. As noted in Sec.~\ref{sec:prelims}, for every individual coupling realization, $50$ such trials are performed, each with an independent initial configuration. These simulations were then repeated for $50$ independent coupling realizations. For each fixed~$N$,  $M_N(t)$ and $q_t(N)$, as defined in~\eqref{eq:def-M} and \eqref{eq:def-q}, are averaged over all $2500$ independent trials. 

\begin{remark}We also tried using the zero-temperature dynamics as a search for the presence of 1-spin-flip stable states for heavy-tailed models, using two methods, dynamic and exhaustive search. In \emph{dynamic search}, we ran independent zero-temperature Glauber dynamics, checking to see whether each absorbing state reached is 1-spin-flip stable (as opposed to $k$-spin-flip stable, with $k>1$). For each type of model, $5\times 10^4$ trials were run in $10$ graph realizations (i.e., $5,000$ trials per realization), resulting in $10$ outcomes. Each outcome records the number of trials in which the system landed in a 1-spin-flip stable state that is not one of the homogenous ground states.  By averaging these data points over the total number of trials, we arrived at the Monte Carlo probability of the system landing in a non-trivial local minimum, for each value of $N$ considered. Consistent with Theorem~\ref{thm:pos-field} and Conjecture~\ref{conj:pos-field}, the dynamic search had probability going to zero rapidly with $N$, of finding any local minima besides the two homogenous ground states. 

In \emph{exhaustive search}, we examined all spin configurations in $\{\pm 1\}^N$ and determined which are 1-spin-flip stable. 
This method obviously can only be used for relatively small systems; we were able to obtain results up to $N=25$, where the probability (with respect to coupling realizations) of finding metastable states appeared to increase with $N$. But these preliminary data were insufficient to draw quantitative conclusions. We defer discussion of existence of local minima (or `traps') in mean-field models to an upcoming paper~\cite{SGNS19}, where we find strong evidence for the existence of many such traps.
\end{remark}

\subsection{Light-tailed distributions}
\label{subsec:light}

In this section we consider the two light-tailed distributions --- randomly diluted using a Bernoulli process and the complete graph with a coupling distribution chosen from a half-normal distribution --- introduced in Sec.~\ref{sec:prelims},  and numerically study their dynamical evolution at zero temperature. In all cases shown, the initial state is chosen by assigning the values of $\pm 1$ to each spin
independently with equal probability, and then allowing the system to
evolve using zero-temperature Glauber dynamics.\footnote{Runs were also conducted in which the magnetization in the initial state was constrained to be exactly zero. In these cases the end results were the same, in that
in each run the system ended in the uniformly positive or negative state; but of course which of the two the final states was reached now depended purely on the dynamical realization.}

\begin{figure}[h]
\centering
\includegraphics[scale=.43]{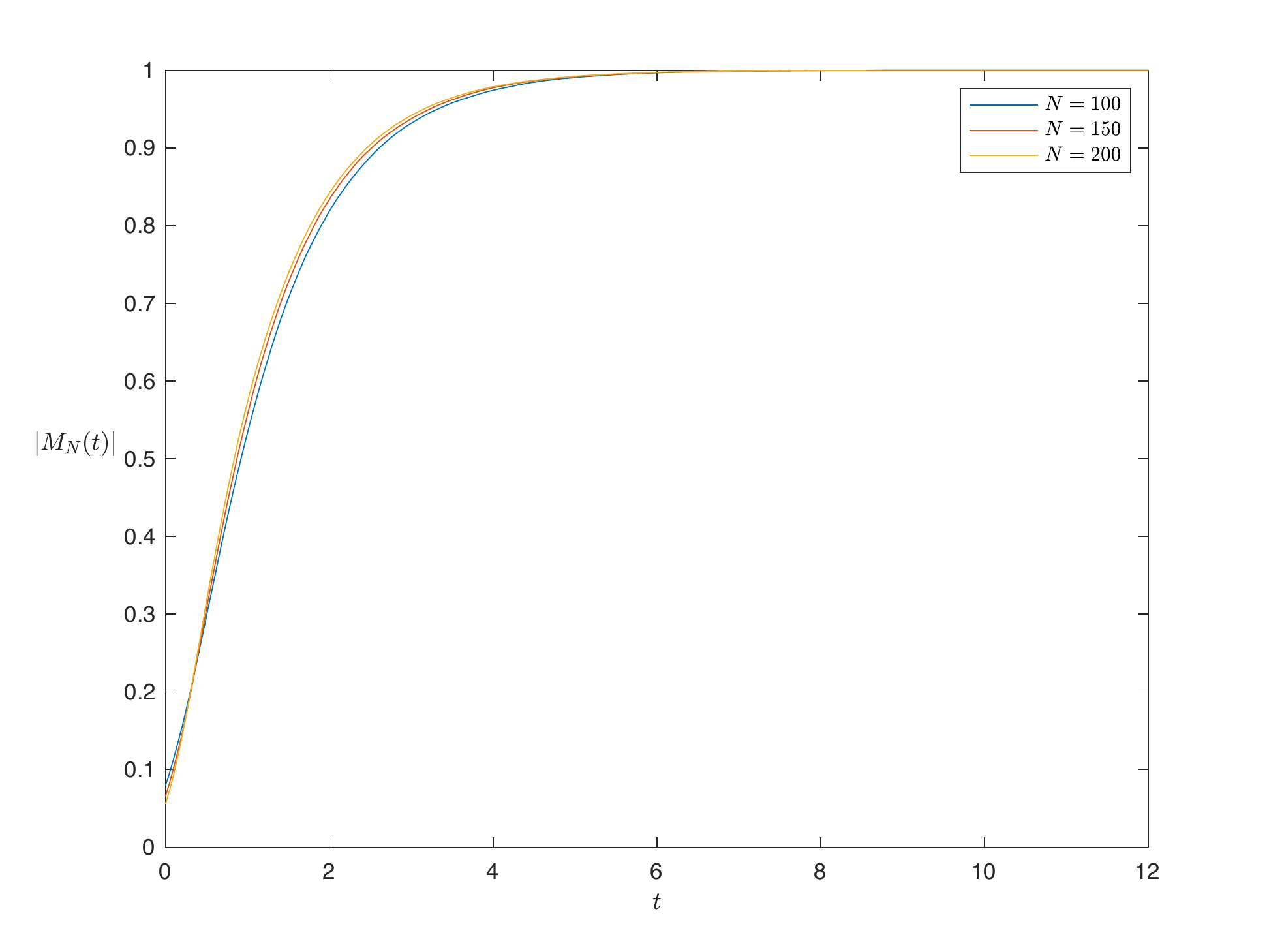}
\hspace{-.7cm}
\includegraphics[scale=.43]{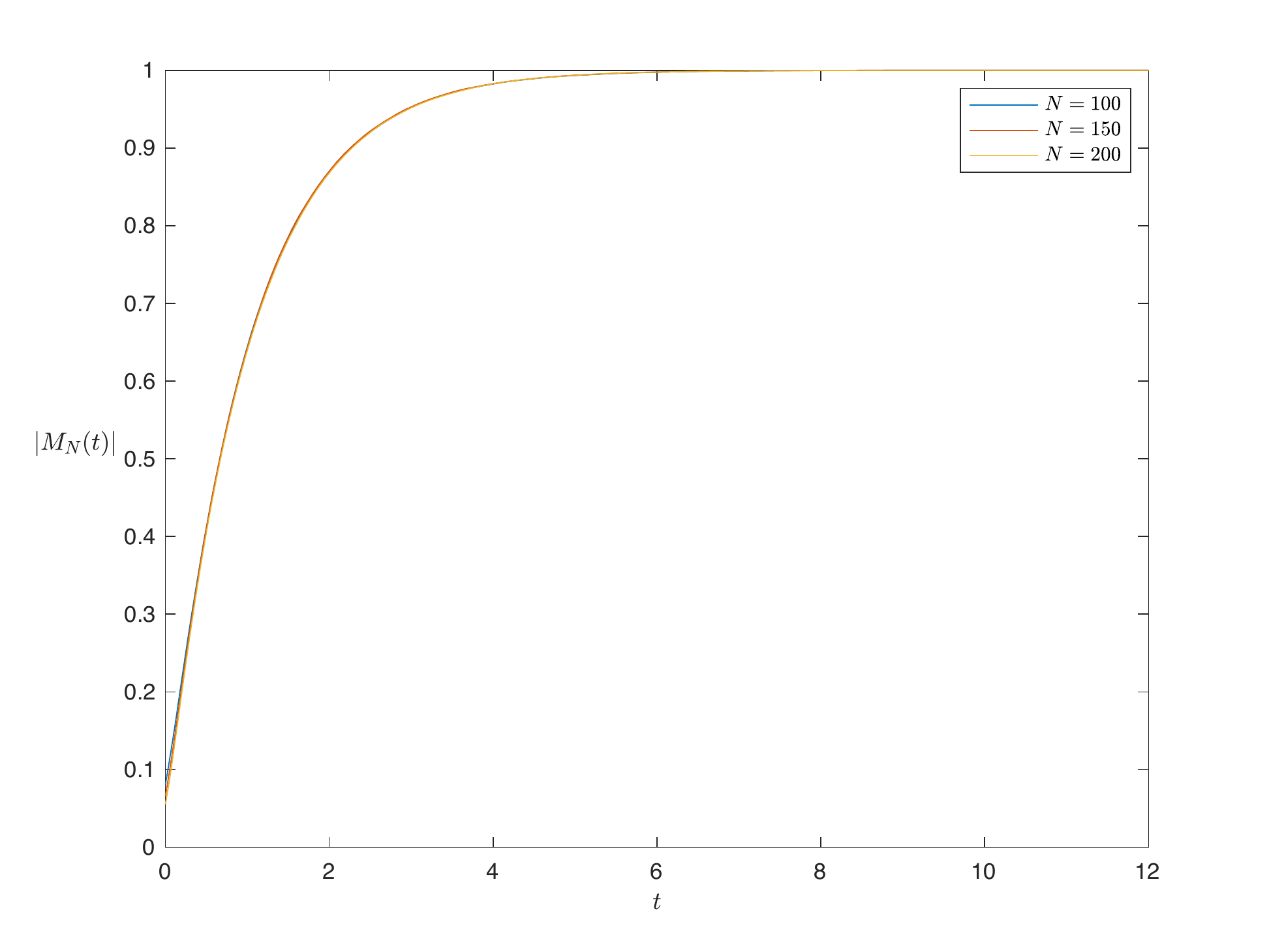}
\hspace{-.7cm}
\caption{$|M_N(t)|$ vs.~time, averaged over 2500 runs in the random ferromagnet with Bernoulli distribution (left), and half-normal distribution (right) for $N=100,150$, and $200$.}
\label{fig:Bernoullimag}
\end{figure}

Fig.~\ref{fig:Bernoullimag} shows the (absolute value of the) magnetization $M_N(t)$ vs.~time for three values of~$N$ ranging from~100 to~200, for both randomly diluted graphs and the complete graph with
a half-normal coupling distribution. According to ~Theorem~1 of~\cite{GNS18} (along with Conjecture~1 in this paper for the half-normal case), as $N\to\infty$ the absolute value of the magnetization in any single run increases monotonically to~1 at $t\to\infty$ for all randomly chosen initial conditions and random zero-temperature dynamical realizations. This is seen in the figure.
In the above and all other figures in this section, the error bars represent the root mean square deviation over all runs for the overlap value at each time step. 

\begin{figure}[h]
\centering
\includegraphics[scale=.43]{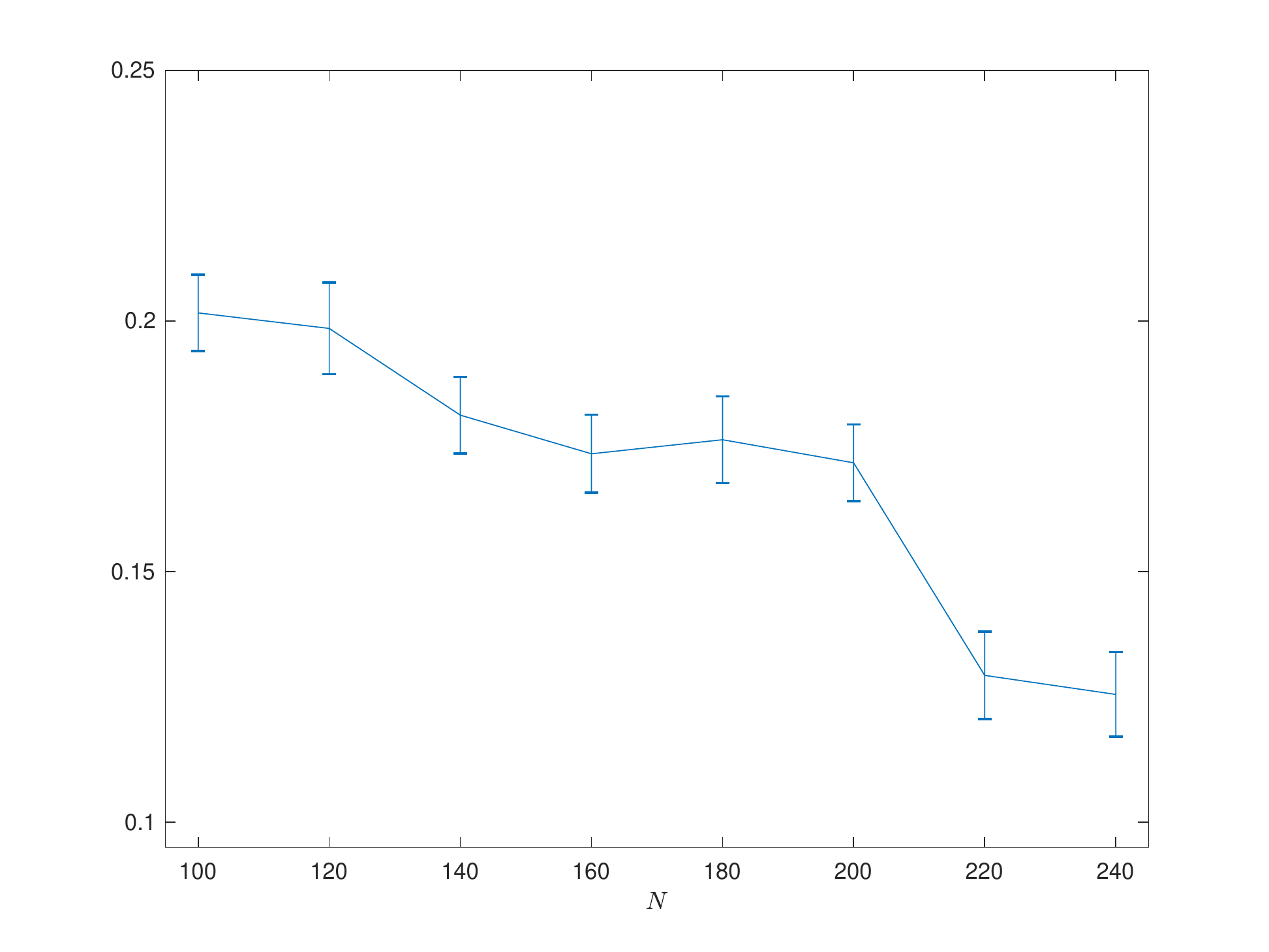}
\hspace{-.7cm}
\includegraphics[scale=.43]{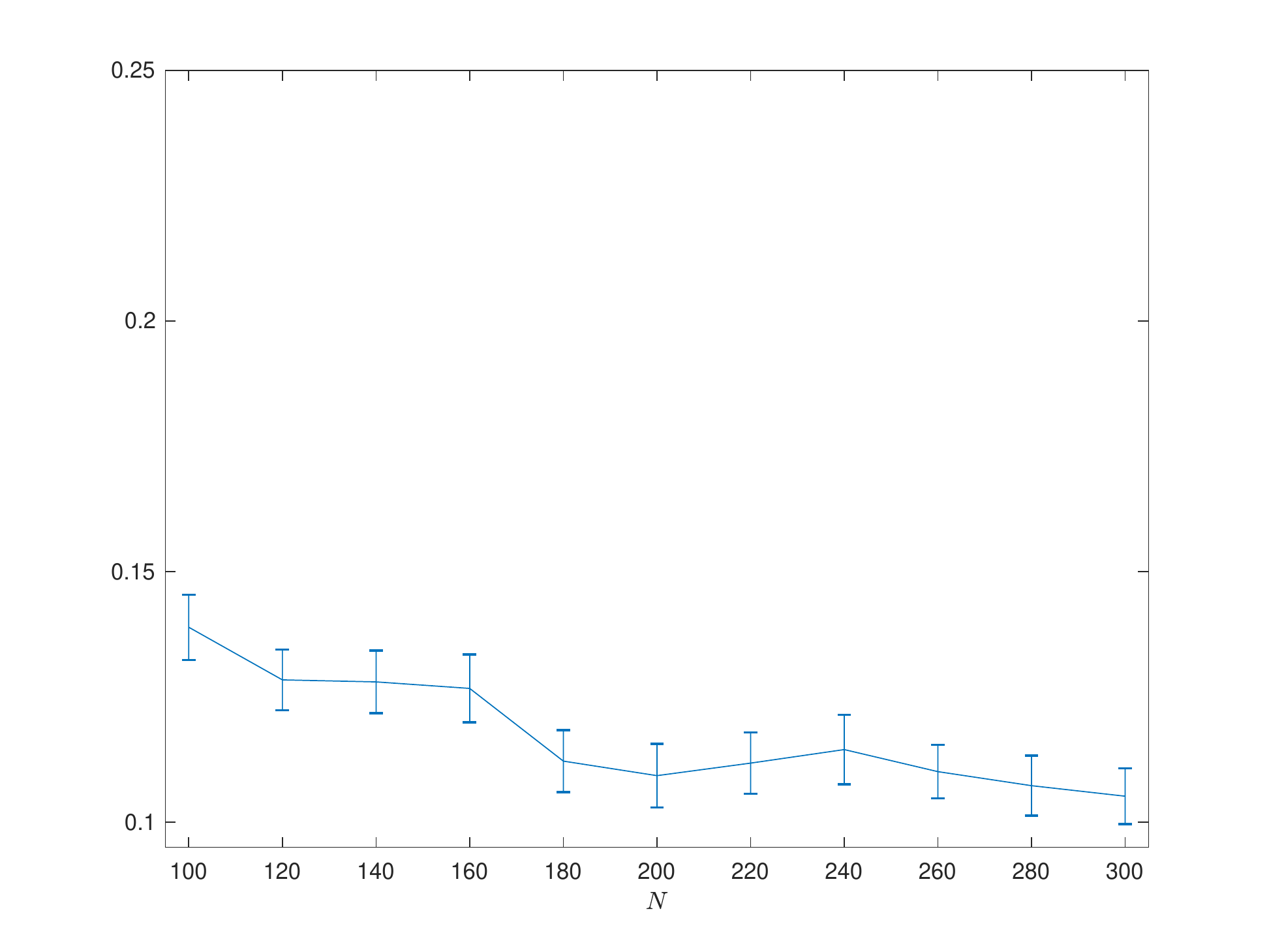}
\hspace{-.7cm}
\caption{Fraction of runs (averaged over graphs, initializations, and dynamics) in which the two twins do not end up in the same homogenous absorbing state (all plus or all minus) vs.\ $N$ in the random ferromagnet with Bernoulli distribution (left) and half-normal distribution (right). Error bars indicate a 95\% confidence interval.}
\label{fig:prop}
\end{figure}

\begin{figure}[h]
\centering
\includegraphics[scale=.43]{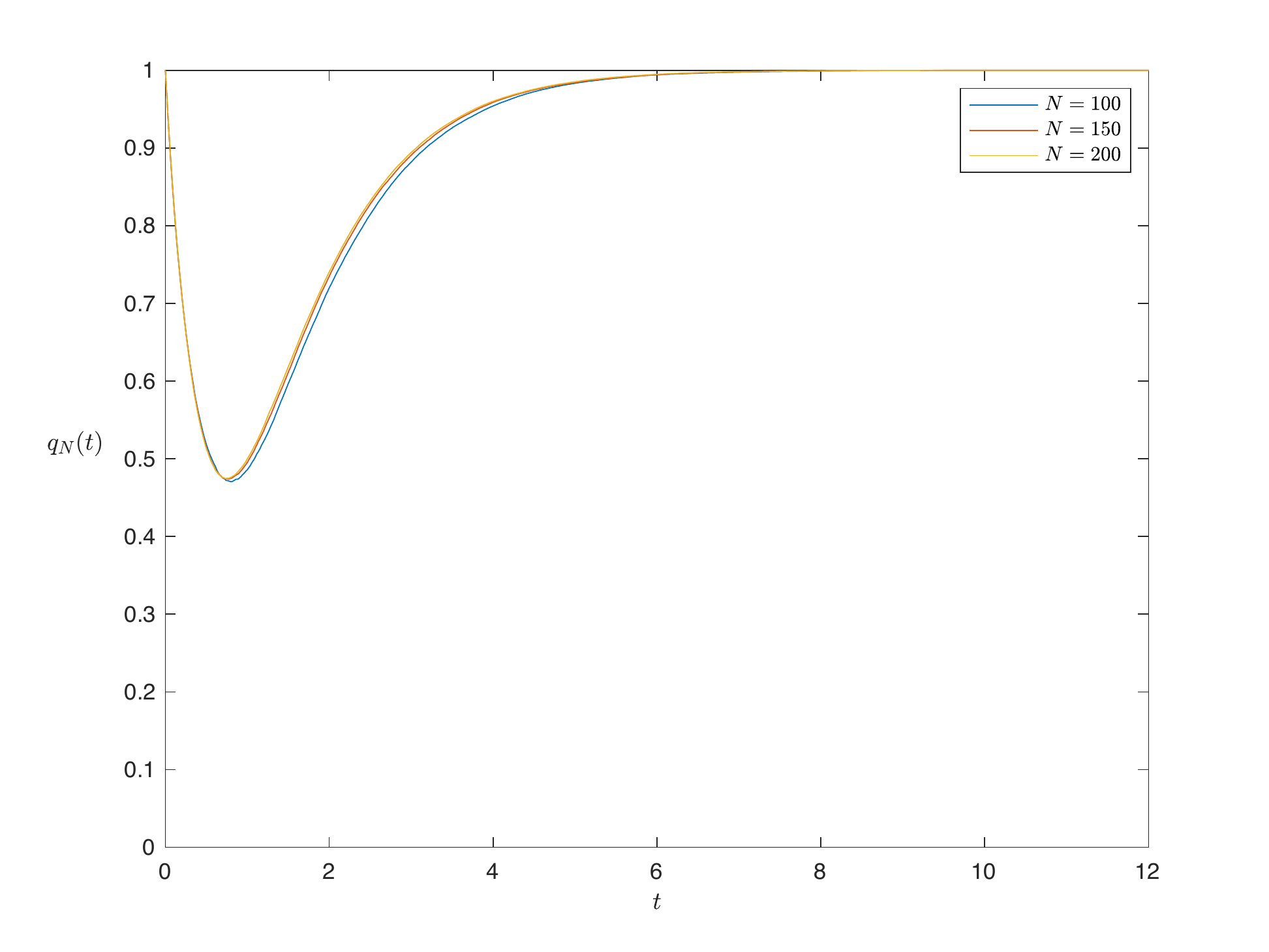}
\hspace{-.7cm}
\includegraphics[scale=.43]{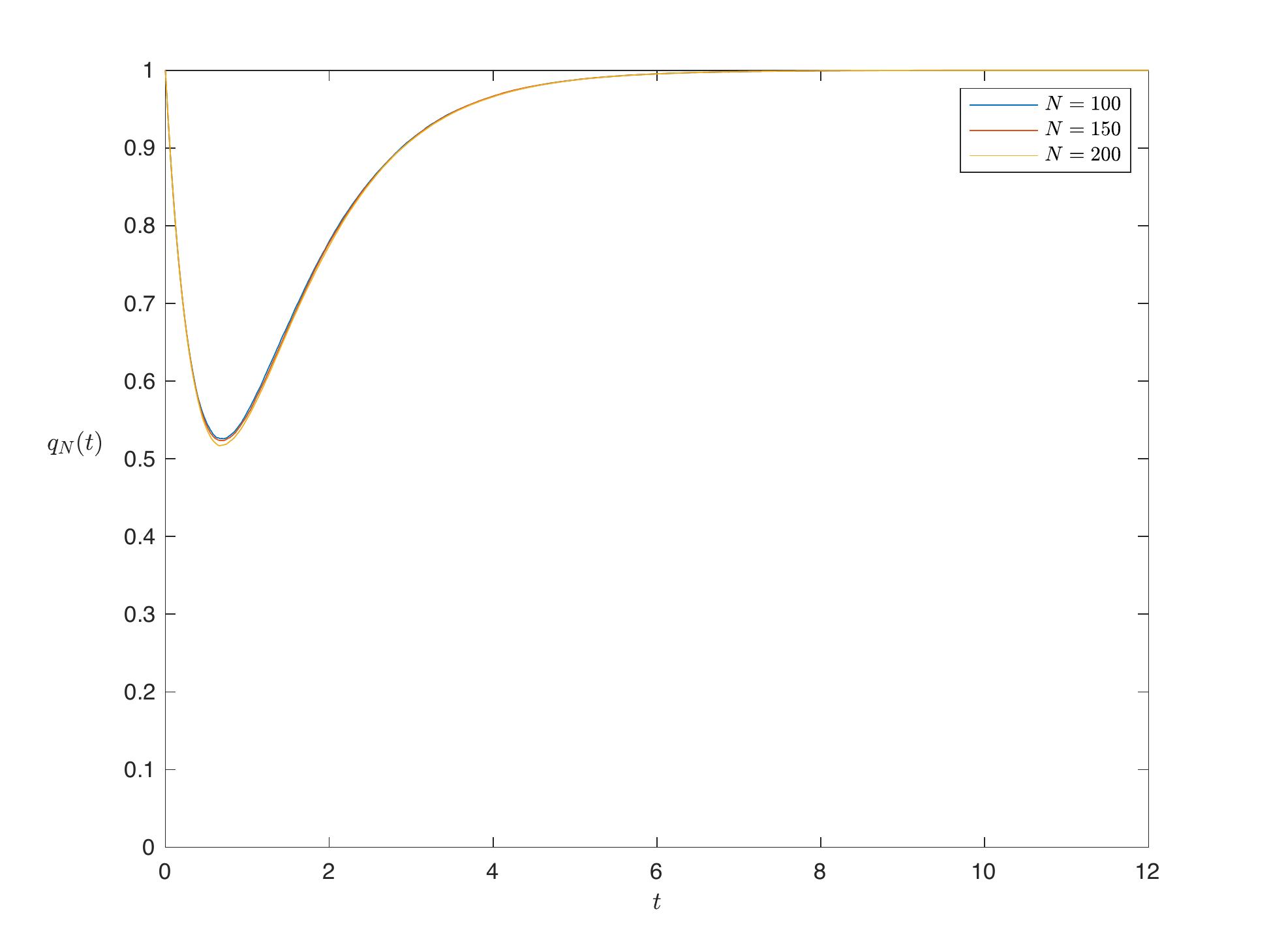}
\hspace{-.7cm}
\caption{$q_N(t)$ vs.~time, averaged over 2500 runs in the random ferromagnet with Bernoulli distribution (left), and half-normal distribution (right) for $N=100,150$, and $200$.}
\label{fig:Bernoulliqt}
\end{figure}

We turn next to the twin overlap, shown in Fig.~\ref{fig:Bernoulliqt} for the same three values of~$N$. Here we have to separate two different effects on the trajectory $q_N(t)$. The first is from situations in which one twin ends in the all $+1$ state and the other in the all $-1$ state; these necessarily contribute $-1$ to the twin overlap and can therefore substantially affect the averaged $q_N(t)$. By Theorem~1 of~\cite{GNS18}, the proportion of runs in which this divergence of the outcome occurs falls to zero as $N\to\infty$ in the presence of Bernoulli couplings, and should fall to zero for any light-tailed distribution.
This conclusion is supported numerically, though the decay rate is quite slow, presumably because with probability of order $N^{-\frac 12^-}$, the initial magnetization is small: $M_N(0)= o(\sqrt N)$. Fig.~\ref{fig:prop} shows the proportion of ``diverging'' runs decreasing as $N$ increases under Bernoulli and half-normal couplings.

The analytical result presented in Eq.~(\ref{eq:qteq}) refers to the $N\to\infty$ limit, in which twin divergences do not occur. Since the finite $N$ effect of these divergences is substantial, an optimal comparison of theory with experiment should consider only those runs in which both twins end up in the same final state. These are shown in the following two figures.\footnote{Because the magnetization calculations consider twins independently, and only the absolute value of the magnetization is of interest, the magnetization results are unaffected by divergences in dynamical outcomes.}
In both cases, at short times $q_N(t)$ falls, reaches a minimum, and then rises to a value approaching one as $N$ increases. The qualitative behavior is easy to understand: the twins begin in the same configuration, but in general different spins are selected at each step in the two independent dynamical realizations and roughly half will flip to lower the energy. Because the selected spins are different with high probability in the two samples, the overlap drops rapidly before reaching a minimum and then rising again (given that in most samples the twins evolve to the same final state).  

Indeed this aligns exactly with the theoretical predictions of Sec.~\ref{sec:theory}, independently of the particulars of the light-tailed distribution. Indeed, we compare $q_N(t)$ for $N=200$ (after removing instances in which the twins end in opposite final states), the largest size studied, and compare its trajectory to the theoretical predictions of~\eqref{eq:result1}. 
\begin{figure}[h]
\centering
\includegraphics[scale=.43]{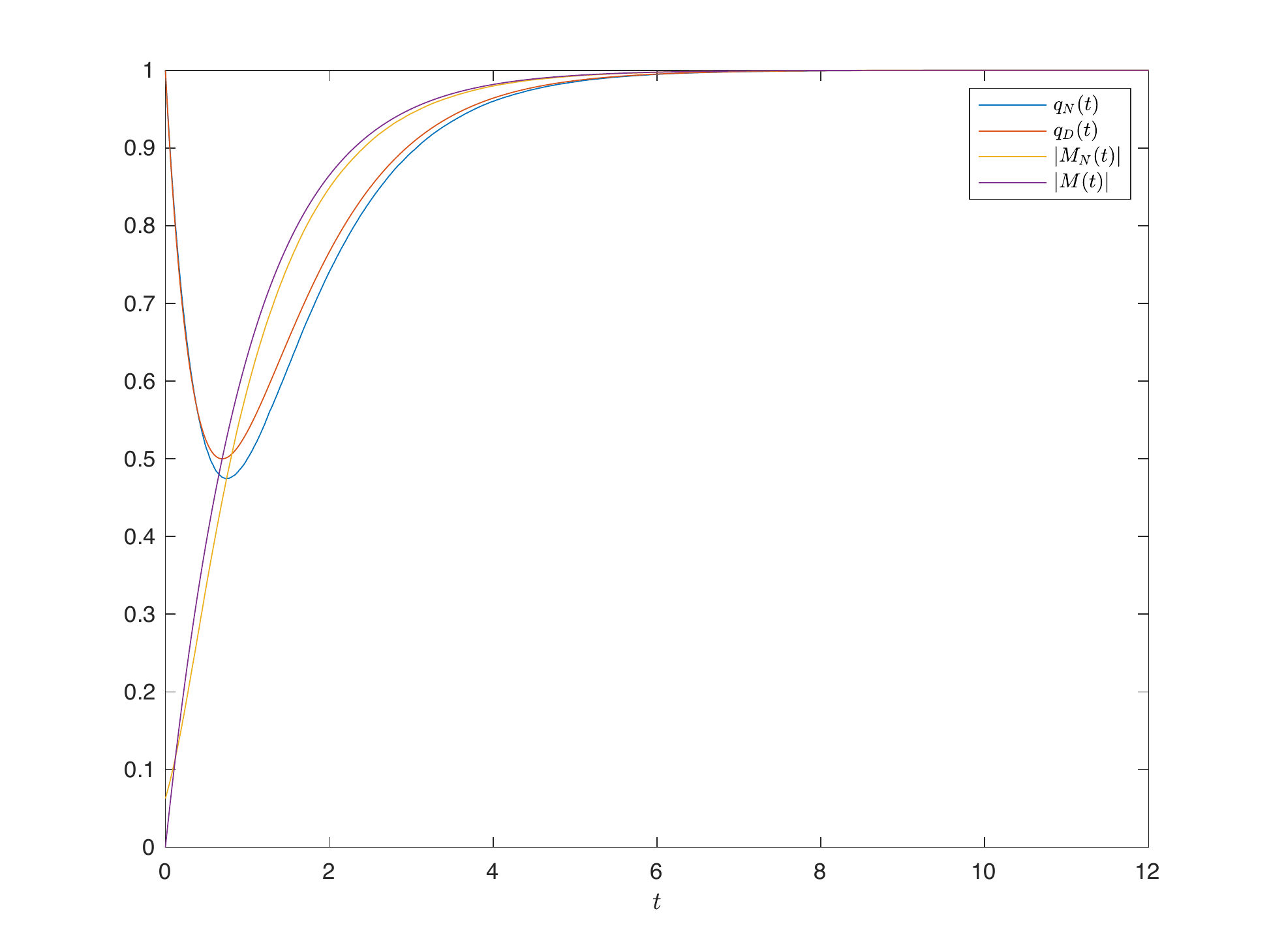}
\hspace{-.7cm}
\includegraphics[scale=.43]{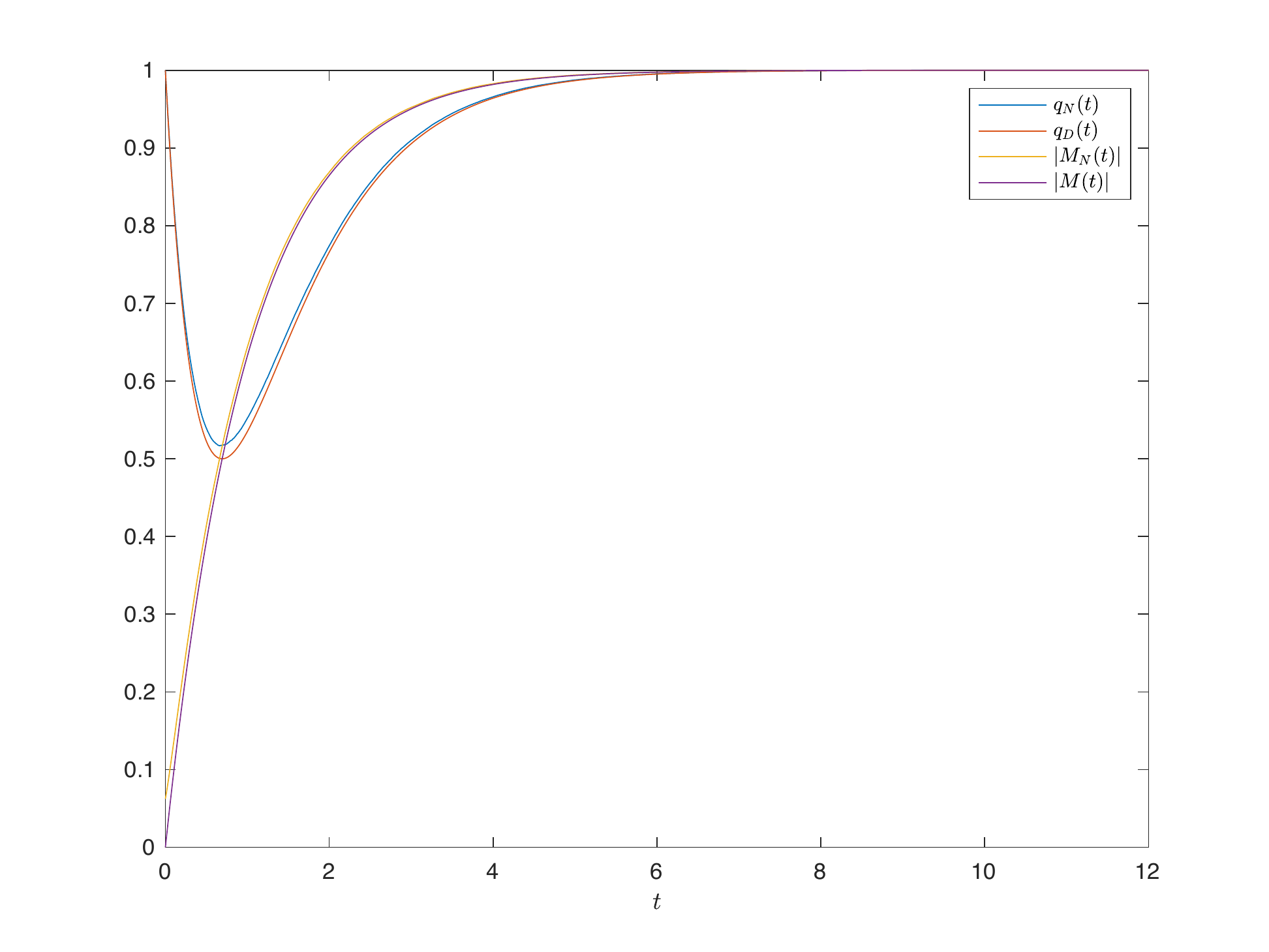}
\hspace{-.7cm}
\caption{A comparison between the numerical pair $|M_N(t)|,q_N(t)$ at the largest $N = 200$ and the theoretical predictions $q_D(t),|M(t)|$ given by~\eqref{eq:result1}--\eqref{eq:mag-theory} in the random ferromagnet with Bernoulli coupling distribution (left) and half-normal distribution (right).}
\label{fig:Bernoulliqt200nodiversions}
\end{figure}
The quantitative values for the minimum overlap and the time at which it's reached are derived in the analytical solution for the dynamical process in Sec.~\ref{sec:theory}; the minimum is predicted to occur at overlap $1/2$ at time $1/e$.
We see from the figures above that with the divergences removed, the agreement between theory and the numerical runs is very good. This suggests that the limiting trajectories $q_N(t)$ and $M_N(t)$ are universal amongst mean-field random ferromagnets with light-tailed coupling distributions, matching the limiting trajectories of the solvable homogenous Curie--Weiss model.

\subsection{Heavy-tailed distributions}
\label{subsec:heavy}

The theory of Sec.~\ref{sec:theory} and the numerics in Section~\ref{subsec:light} pertained to mean-field random ferromagnets with
light-tailed distributions, such as  Bernoulli or the half-normal, whose tails are sub-exponential or sub-Gaussian, say. In principle, we expect all such models to have
$q_\infty(N)\to 1$ as $N\to\infty$, with the magnetization in a given run tending
to $\pm 1$ as $t\to\infty$; moreover, we expect the sign of the magnetization to be asymptotically, fully determined, by the initial bias.
As noted earlier, this result has been proven for the randomly diluted
infinite-range ferromagnet~\cite{GNS18}. An implication is that
in light-tailed models the normalized volume of the union of the
domains of attraction of (non-trivial) 1-spin-flip stable states tends to zero as $N\to\infty$.

As a contrast, it is useful to examine models with so-called heavy-tailed
distributions, having polynomial tails for some sufficiently small exponent, say small enough that the distribution has no first moment. In~\cite{GNS18} a very general class of such distributions was treated including Pareto distributions with parameter $\alpha\in (0,1)$; the half-Cauchy distribution is a boundary case where $\alpha=1$. 
For these coupling distributions, we expect 
behavior different from that found for light-tailed distributions. It is a consequence of rigorous results in~\cite{GNS18} that with strictly positive probability, there are exponentially many in $N$ 1-spin flip stable states, and the magnetization $\lim_{t\to\infty}M(t)$ stays bounded away from $\pm 1$ as $N\to\infty$; moreover, the value of $\lim_{t\to\infty}q_N(t)$ stays bounded away from $0$ and $1$ as $N\to\infty$. These are inferred from the presence of order $N$ many so-called bully bonds {(whose existence is ensured by the classical fact that the maximum of $N$ i.i.d.\  positive heavy-tailed random variables is on the same order as their sum.)}

One may then be interested in the trajectory to absorption in these cases where the randomness of the dynamics plays a significant role in determining the absorbing state. We investigate this numerically in this section. Our results showed qualitatively similar behaviors between several choices of heavy-tailed distributions---Pareto for $\alpha =\frac 12$ and half-Cauchy---so we present only the numerics for the edge-case of the half-Cauchy distribution. {N.B.\ while the behaviors of different heavy-tailed distributions were qualitatively similar, unlike the light-tailed case they were not quantitatively aligned; indeed we do not expect any universality in the limiting trajectories here.} As before, the initial state is chosen by assigning the values of
$\pm 1$ to each spin independently with equal probability, and then the system is allowed to evolve using zero-temperature Glauber dynamics.

\begin{figure}[h]
\centering
\includegraphics[scale=.43]{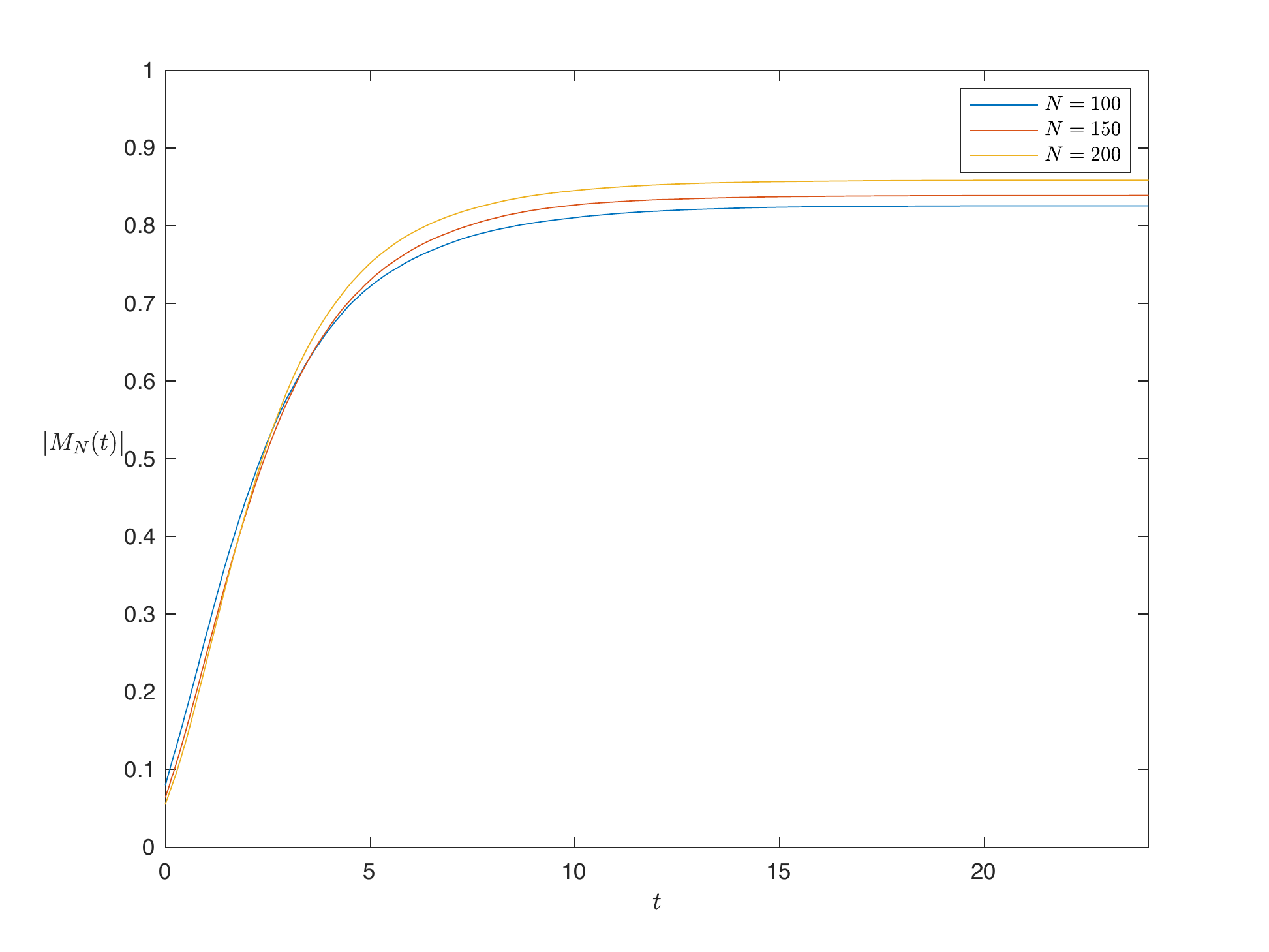}
\hspace{-.7cm}
\includegraphics[scale=.43]{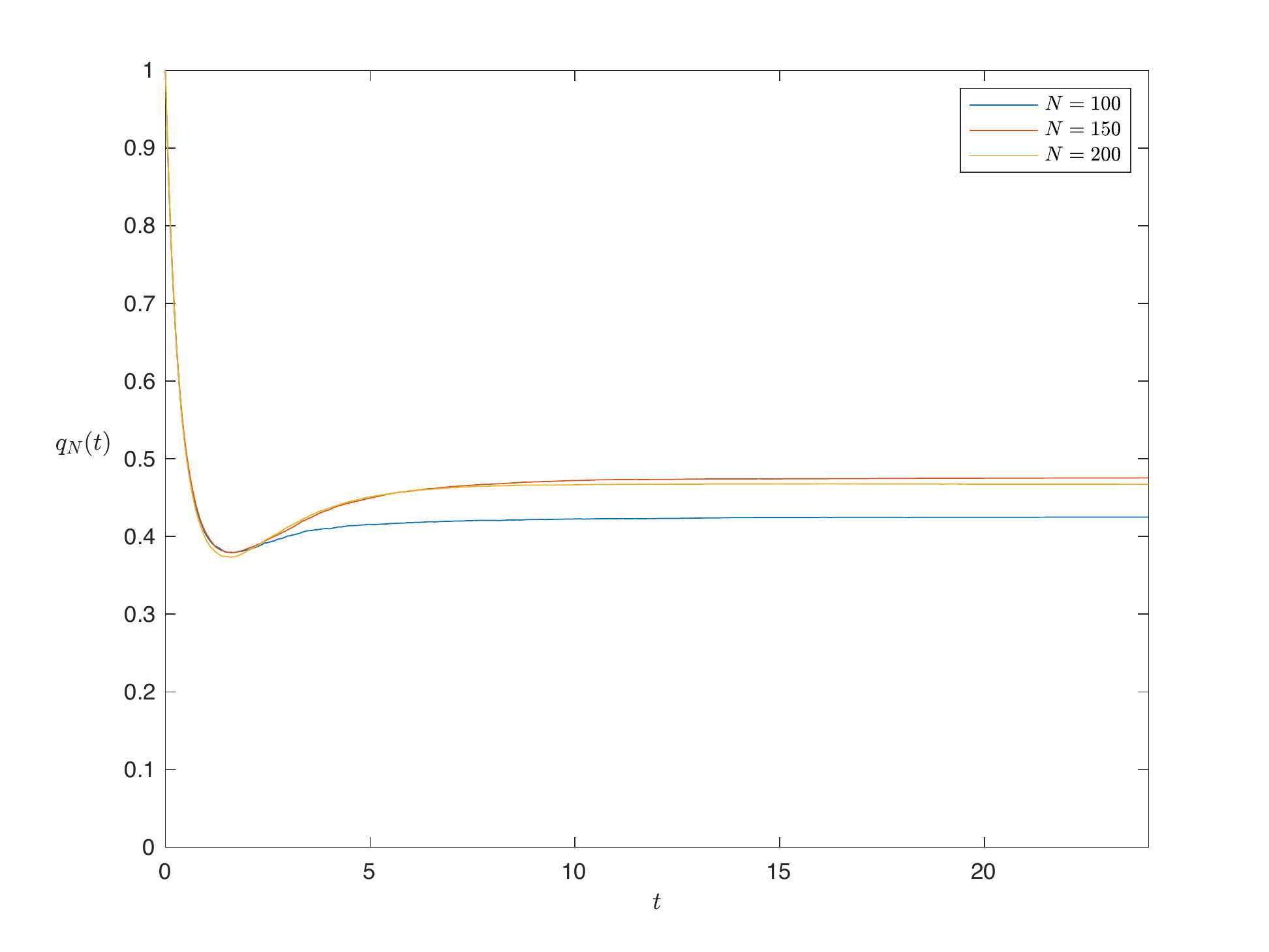}
\hspace{-.7cm}
\caption{$|M_N(t)|$ (left) and $q_N(t)$ (right) vs time, in a mean-field random ferromagnet with half-Cauchy distributed couplings for $N= 100,150,200$, averaged over 2500 independent realizations of the graph, initialization, and dynamics. }
\label{fig:Cauchymag}
\end{figure}

Fig.~\ref{fig:Cauchymag} shows the magnetization $M_N(t)$ vs.~time for various system sizes ranging from $N=100$ to $N=200$, from time zero until beyond the absorption time of the chain; in particular, the magnetization has converged to its final $t\to\infty$ value.  One
sees from the figure that the (absolute value of) the final magnetization
is approximately~.5. (The initial value in all cases is noticeably greater
than zero because of the relatively small sizes of the samples; however,
one can see the initial value falling to zero as $N^{-1/2}$.) This suggests that there is a dominant ferromagnetic effect that is competing against the glassy fluctuations induced by the strong couplings (e.g., by bully bonds starting in alignment).

This indicates that, unlike the light-tailed case, when the couplings are heavy-tailed, with positive probability the dynamical realizations end up in 1 spin-flip stable states besides the all-plus and all-minus configurations.   
In fact, further numerics (not depicted here) found that the probability of a dynamical run ``landing'' in a non-homogenous 1 spin-flip stable state 
approaches one as $N$ gets large, so that typical absorbing states have partial, but not full magnetization. Namely, it appeared that the typical value of $\lim_{t\to\infty} M_N(t)$ aligned well with its average value depicted in Fig.~\ref{fig:Cauchymag}.

The time evolution of the twin overlap $q_N(t)$ for the same three system sizes of $N = 100, 150, 200$ is
also shown in Fig.~\ref{fig:Cauchymag}. Although the curves qualitatively appear
similar to those for the light-tailed cases, crucially the limit as $t,N\to\infty$ of $q_N(t)$ remains bounded away from zero and one. Though it is unclear what number the twin overlap is converging to, it seems it is somewhere in $[.5,1)$; the minimum twin overlap rapidly decays to strictly below 0.5, before attaining its absorbing value. {We do not have any reason to believe $\lim_{N\to\infty} \lim_{t\to\infty} q_N(t)$ is converging to any distribution-independent number like .5 (as it does for the 1D random ferromagnet). The choice of heavy-tailed distribution plays a significant role in determining the geometry of the influence graph given by the coupling realization. Moreover, the numerical values of $\lim_{t\to\infty} q_N(t)$ may still have large finite-$N$ effects, including from partial divergences wherein one twin ends up with magnetization, say, $+.85$, while the other ends up with magnetization $-.85$.}

We can deduce that, unlike the light-tailed cases, the randomness of the dynamical evolution plays a significant role in
determining the final state (though still smaller than that played by the
initial condition). However, we emphasize the qualitative similarities in the profile of $q(t)$ as it approaches its $t\to\infty$ limit, between the heavy-tailed and light-tailed cases. 

\subsection{Other geometries: the 1D random ferromagnet}\label{subsec:1D-numerics}
The first discussion in the theoretical derivations of Section~\ref{sec:theory} pertained to systems in which the only spin flips were from minus to plus; due to the rigorous results of~\cite{GNS18}, this is indeed the case after  $o(N)$ time steps in the infinite-range ferromagnets with light tails. In Section~\ref{sec:tree}, we derived solutions for more complicated situations which we termed ``highly disordered" as a starting point for deriving the approach to equilibrium in general systems; these included as a particular example the 1D random ferromagnet with general coupling distribution. 

\begin{figure}[h]
\centering
\includegraphics[scale=.57]{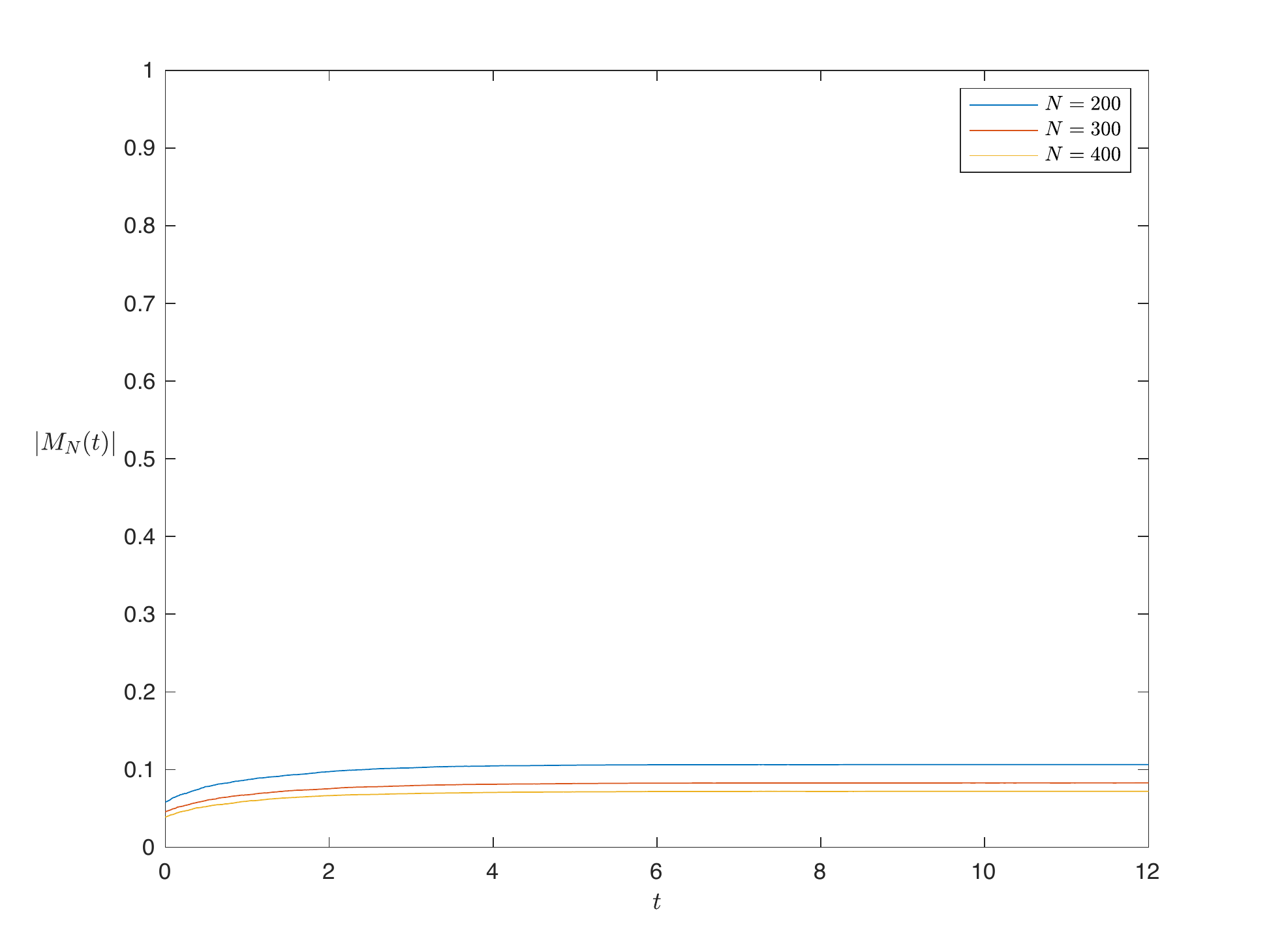}
\caption{$|M_N(t)|$ vs time, averaged over 2500 independent realizations of the graph, initialization, and dynamics, in a 1D random ferromagnet with half-normal couplings, for $N= 200,300,400$.}
\label{fig:1D-mag}
\end{figure}

\begin{figure}[h]
\centering
\includegraphics[scale=.43]{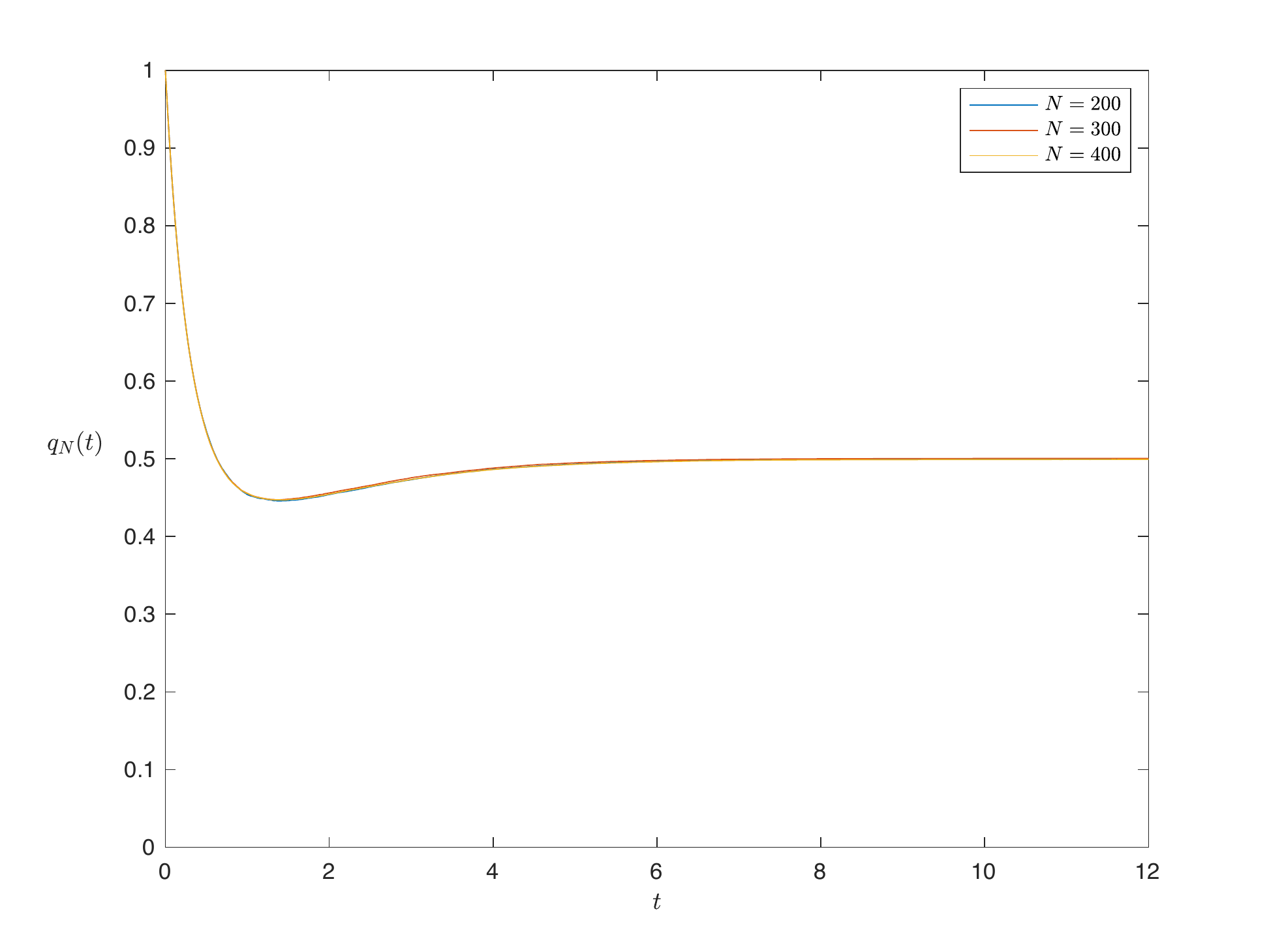}
\hspace{-.7cm}
\includegraphics[scale=.43]{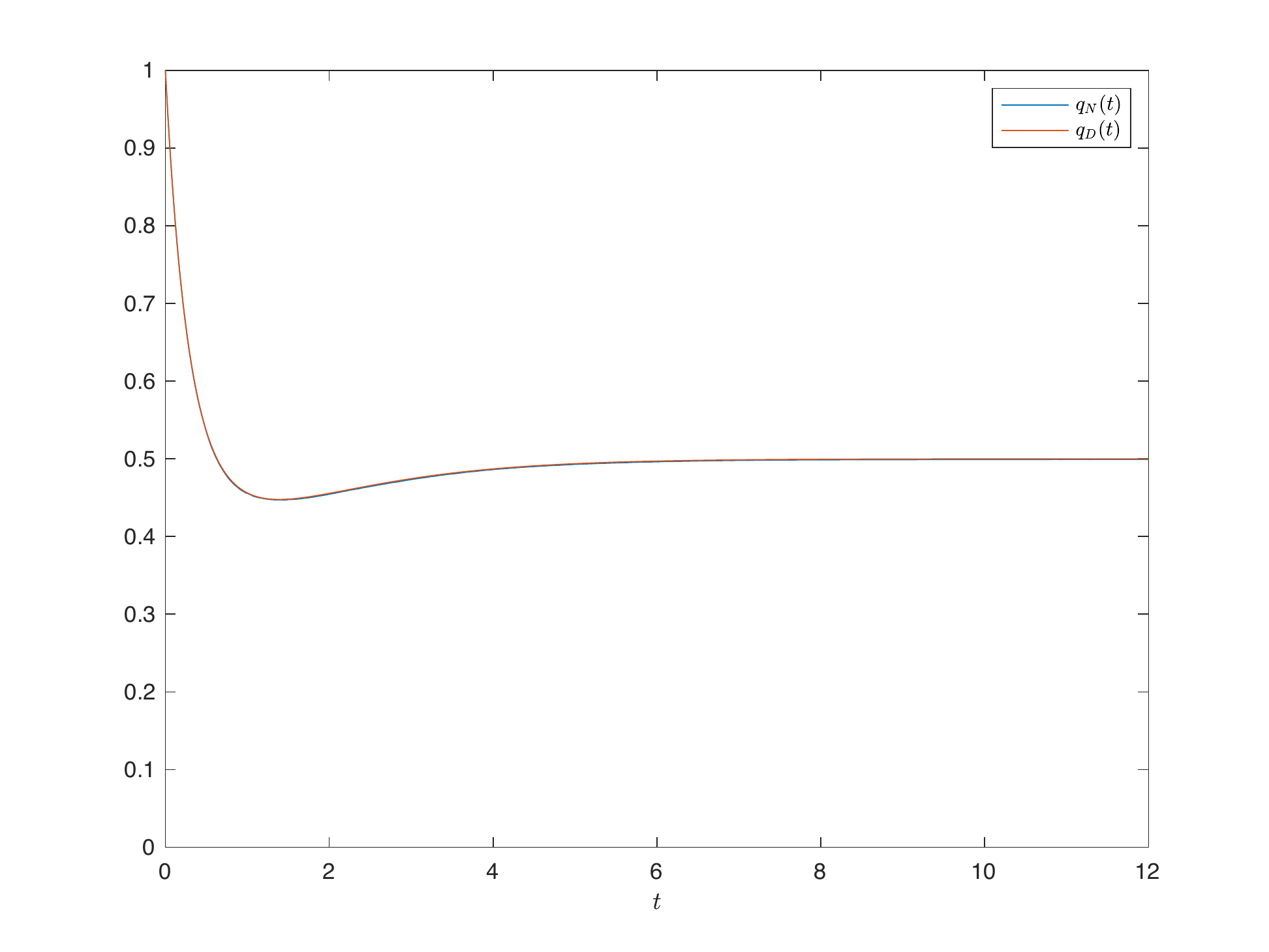}
\hspace{-.7cm}
\caption{Left: $q_N(t)$ vs time in a 1D random ferromagnet with half-normal couplings, for $N= 200,300,400$, averaged over 2500 independent realizations of the graph, initialization, and dynamics. Right: the $N=400$ curve is plotted against the theoretical solution from~\eqref{eq:1D-theory}.}
\label{fig:1D-qt}
\end{figure}

The numerics we present in this section are for the case of the 1D random ferromagnet, where the theoretical solution is also tractable, for comparison. 
It was shown in~\cite{NNS00} that a 1D random ferromagnet with a non-degenerate coupling distribution that has  a density with respect to Lebesgue, has $\lim_{t\to\infty}q_D(t) = \frac 12$. As seen in Section~\ref{sec:theory}, since the space breaks up easily into segments of finite length between bully bonds and, what were called \emph{weak links} in~\cite{NNS00}, for a fixed realization of the coupling, averaged over the initial configuration and the dynamical evolution, the final configuration consists of i.i.d.\ assignments of plus and minus spins to fixed segments of order one length.   

As such on average, while the magnetization at time zero is given by a sum of $N$ i.i.d.\ plus and minus spins, in the $t\to\infty$ limit, it is on average given by a weighted sum of $a_\nu N \leq N$ i.i.d.\ plus and minus spins, where the respective weights and $a_\nu$ are given by the coupling realization---thus the variance only increases, and one would expect the absolute magnetization to increase in time (though it will remain $O(N^{-1/2})$). Indeed Figure~\ref{fig:1D-mag} demonstrates that this is the case. By this reasoning, it is clear that the limiting trajectory $M(t): = \lim_{N\to\infty} M_N(t)$ will be constant at zero.

As a consequence of this geometric structure, as we saw in Section~\ref{sec:theory}, the time evolution of $q_D(t)$ also admits an exact solution. In Figure~\ref{fig:1D-qt}, we see have plotted the time evolution of the overlap for $N= 200,300,400$ in Figure~\ref{fig:1D-qt}, left. Interestingly, we again see the same qualitative approach to equilibrium as in the mean-field models, wherein the twins initially experience a rapid divergence, dropping to some strictly positive minimal overlap before converging to their final $\lim_{t\to\infty} q_D(t)$, which in this case is~$\frac 12$. As expected, as $N$ diverges, the trajectory of $q_D(t)$ converges to its predicted trajectory given in~\eqref{eq:1D-theory}: see Figure~\ref{fig:1D-qt}, right. One finds that the rate of convergence as $N\to\infty$ to the predicted limiting trajectory is much faster in this low-dimensional setup than in both the heavy-tailed and light-tailed random ferromagnets.  

\medskip \noindent
\textbf{Acknowledgments.} We thank the anonymous referee for helpful suggestions. The research of CMN was supported in part by
U.S. NSF Grant DMS-1507019. DLS thanks the Aspen Center for Physics, which is supported by National Science Foundation grant PHY-1607611.

\bibliographystyle{abbrv}
\bibliography{refs}

\end{document}